\documentclass[12pt]{article}
\usepackage[dvips]{graphicx}
\usepackage{epsfig}
\usepackage{latexsym,amsmath,amsfonts,amssymb}
\usepackage[latin1]{inputenc}
\usepackage[american]{babel}
\usepackage[dvips]{graphicx}
\usepackage{bbm}
\usepackage[unicode]{hyperref}
\def\im{Invent. Math.}

\def\a{\alpha}
\def\b{\beta}
\def\c{\gamma}
\def\d{\delta}
\def\f{\phi}               
\def\vf{\varphi}  \def\tvf{\tilde{\varphi}}
\def\vp{\varphi}
\def\g{\gamma}
\def\h{\eta}
\def\j{\psi}
\def\k{\kappa}                    
\def\l{\lambda}
\def\m{\mu}
\def\n{\nu}
\def\o{\omega}  \def\w{\omega}

\def\q{\theta}  \def\th{\theta}                  
\def\r{\rho}                                     
\def\s{\sigma}                                   
\def\t{\tau}
\def\u{\upsilon}
\def\x{\xi}
\def\z{\zeta}
\def\pt{\tilde{\varphi}}
\def\tt{\tilde{\theta}}
\def\lab{\label}
\def\6{\partial}
\def\wg{\wedge}
\def\bpsi{\bar{\psi}}
\def\bt{\bar{\theta}}
\def\bvf{\bar{\varphi}}

\DeclareMathOperator{\tr}{tr}

\newcommand{\be}{\begin{equation}}
\newcommand{\ee}{\end{equation}}
\newcommand{\beq}{\begin{equation}}
\newcommand{\eeq}{\end{equation}}
\newcommand{\bea}{\begin{eqnarray}}
\newcommand{\eea}{\end{eqnarray}}

\newcommand{\ba}{\begin{eqnarray}}
\newcommand{\ea}{\end{eqnarray}}
\def\NO{\nonumber}
\newcommand{\beqs}{\begin{eqnarray}}
\newcommand{\eeqs}{\end{eqnarray}}
\newcommand{\bal}{\begin{aligned}}
\newcommand{\eal}{\end{aligned}}

\begin{document}
\baselineskip=15.5pt
\pagestyle{plain}
\setcounter{page}{1}


\def\del{{\partial}}
\def\vev#1{\left\langle #1 \right\rangle}
\def\cn{{\cal N}}
\def\co{{\cal O}}
\def\IC{{\mathbb C}}
\def\IR{{\mathbb R}}
\def\IZ{{\mathbb Z}}
\def\RP{{\bf RP}}
\def\CP{{\bf CP}}
\def\Poincare{{Poincar\'e }}
\def\tr{{\rm tr}}
\def\tp{{\tilde \Phi}}

\def\TL{\hfil$\displaystyle{##}$}
\def\TR{$\displaystyle{{}##}$\hfil}
\def\TC{\hfil$\displaystyle{##}$\hfil}
\def\TT{\hbox{##}}
\def\HLINE{\noalign{\vskip1\jot}\hline\noalign{\vskip1\jot}}
\def\seqalign#1#2{\vcenter{\openup1\jot
   \halign{\strut #1\cr #2 \cr}}}
\def\lbldef#1#2{\expandafter\gdef\csname #1\endcsname {#2}}
\def\eqn#1#2{\lbldef{#1}{(\ref{#1})}%
\begin{equation} #2 \label{#1} \end{equation}}
\def\eqalign#1{\vcenter{\openup1\jot
     \halign{\strut\span\TL & \span\TR\cr #1 \cr
    }}}
\def\eno#1{(\ref{#1})}
\def\href#1#2{#2}
\def\half{\frac{1}{2}}

\def\ads{{\it AdS}}
\def\adsp{{\it AdS}$_{p+2}$}
\def\cft{{\it CFT}}

\newcommand{\ber}{\begin{eqnarray}}
\newcommand{\eer}{\end{eqnarray}}

\newcommand{\beqar}{\begin{eqnarray}}
\newcommand{\cN}{{\cal N}}
\newcommand{\cO}{{\cal O}}
\newcommand{\cA}{{\cal A}}
\newcommand{\cT}{{\cal T}}
\newcommand{\cF}{{\cal F}}
\newcommand{\cC}{{\cal C}}
\newcommand{\cR}{{\cal R}}
\newcommand{\cW}{{\cal W}}
\newcommand{\eeqar}{\end{eqnarray}}
\newcommand{\tht}{\thteta}
\newcommand{\lm}{\lambda}\newcommand{\Lm}{\Lambda}


\newcommand{\nonu}{\nonumber}
\newcommand{\oh}{\displaystyle{\frac{1}{2}}}
\newcommand{\dsl}
   {\kern.06em\hbox{\raise.15ex\hbox{$/$}\kern-.56em\hbox{$\partial$}}}
\newcommand{\id}{i\!\!\not\!\partial}
\newcommand{\as}{\not\!\! A}
\newcommand{\ps}{\not\! p}
\newcommand{\ks}{\not\! k}
\newcommand{\D}{{\cal{D}}}
\newcommand{\dv}{d^2x}
\newcommand{\Z}{{\cal Z}}
\newcommand{\N}{{\cal N}}
\newcommand{\Dsl}{\not\!\! D}
\newcommand{\Bsl}{\not\!\! B}
\newcommand{\Psl}{\not\!\! P}
\newcommand{\eeqarr}{\end{eqnarray}}
\newcommand{\ZZ}{{\rm \kern 0.275em Z \kern -0.92em Z}\;}


\def\del{{\delta^{\hbox{\sevenrm B}}}} \def\ex{{\hbox{\rm e}}}
\def\azb{A_{\bar z}} \def\az{A_z} \def\bzb{B_{\bar z}} \def\bz{B_z}
\def\czb{C_{\bar z}} \def\cz{C_z} \def\dzb{D_{\bar z}} \def\dz{D_z}
\def\im{{\hbox{\rm Im}}} \def\mod{{\hbox{\rm mod}}} \def\tr{{\hbox{\rm Tr}}}
\def\ch{{\hbox{\rm ch}}} \def\imp{{\hbox{\sevenrm Im}}}
\def\trp{{\hbox{\sevenrm Tr}}} \def\vol{{\hbox{\rm Vol}}}
\def\rl{\Lambda_{\hbox{\sevenrm R}}} \def\wl{\Lambda_{\hbox{\sevenrm W}}}
\def\fc{{\cal F}_{k+\cox}} \def\vev{vacuum expectation value}
\def\nodiv{\mid{\hbox{\hskip-7.8pt/}}}
\def\ie{{\em i.e.}}
\def\ie{\hbox{\it i.e.}}

\def\CC{{\mathchoice
{\rm C\mkern-8mu\vrule height1.45ex depth-.05ex
width.05em\mkern9mu\kern-.05em}
{\rm C\mkern-8mu\vrule height1.45ex depth-.05ex
width.05em\mkern9mu\kern-.05em}
{\rm C\mkern-8mu\vrule height1ex depth-.07ex
width.035em\mkern9mu\kern-.035em}
{\rm C\mkern-8mu\vrule height.65ex depth-.1ex
width.025em\mkern8mu\kern-.025em}}}

\def\RR{{\rm I\kern-1.6pt {\rm R}}}
\def\NN{{\rm I\!N}}
\def\ZZ{{\rm Z}\kern-3.8pt {\rm Z} \kern2pt}
\def\IB{\relax{\rm I\kern-.18em B}}
\def\ID{\relax{\rm I\kern-.18em D}}
\def\II{\relax{\rm I\kern-.18em I}}
\def\IP{\relax{\rm I\kern-.18em P}}
\newcommand{\CS}{{\scriptstyle {\rm CS}}}
\newcommand{\CSs}{{\scriptscriptstyle {\rm CS}}}
\newcommand{\rc}{\nonumber\\}
\newcommand{\bear}{\begin{eqnarray}}
\newcommand{\eear}{\end{eqnarray}}

\newcommand{\LL}{{\cal L}}

\def\mani{{\cal M}}
\def\calo{{\cal O}}
\def\calb{{\cal B}}
\def\calw{{\cal W}}
\def\calz{{\cal Z}}
\def\cald{{\cal D}}
\def\calc{{\cal C}}
\def\to{\rightarrow}
\def\ele{{\hbox{\sevenrm L}}}
\def\ere{{\hbox{\sevenrm R}}}
\def\zb{{\bar z}}
\def\wb{{\bar w}}
\def\nodiv{\mid{\hbox{\hskip-7.8pt/}}}
\def\menos{\hbox{\hskip-2.9pt}}
\def\dr{\dot R_}
\def\drr{\dot r_}
\def\ds{\dot s_}
\def\da{\dot A_}
\def\dga{\dot \gamma_}
\def\ga{\gamma_}
\def\dal{\dot\alpha_}
\def\al{\alpha_}
\def\cl{{closed}}
\def\cls{{closing}}
\def\vev{vacuum expectation value}
\def\tr{{\rm Tr}}
\def\to{\rightarrow}
\def\too{\longrightarrow}


\def\a{\alpha}
\def\b{\beta}
\def\c{\gamma}
\def\d{\delta}
\def\e{\epsilon}           
\def\F{\Phi}
\def\f{\phi}               
\def\vf{\varphi}  \def\tvf{\tilde{\varphi}}
\def\vp{\varphi}
\def\g{\gamma}
\def\h{\eta}
\def\j{\psi}
\def\k{\kappa}                    
\def\l{\lambda}
\def\m{\mu}
\def\n{\nu}
\def\o{\omega}  \def\w{\omega}
\def\q{\theta}  \def\th{\theta}                  
\def\r{\rho}                                     
\def\s{\sigma}                                   
\def\t{\tau}
\def\u{\upsilon}
\def\x{\xi}
\def\X{\Xi}
\def\z{\zeta}
\def\pt{\tilde{\varphi}}
\def\tt{\tilde{\theta}}
\def\lab{\label}
\def\6{\partial}
\def\wg{\wedge}
\def\atanh{{\rm arctanh}}
\def\bpsi{\bar{\psi}}
\def\bt{\bar{\theta}}
\def\bvf{\bar{\varphi}}

%

\newfont{\namefont}{cmr10}
\newfont{\addfont}{cmti7 scaled 1440}
\newfont{\boldmathfont}{cmbx10}
\newfont{\headfontb}{cmbx10 scaled 1728}
\newcommand{\re}{\,\mathbb{R}\mbox{e}\,}
\newcommand{\hyph}[1]{$#1$\nobreakdash-\hspace{0pt}}
\providecommand{\abs}[1]{\lvert#1\rvert}
\newcommand{\Nugual}[1]{$\mathcal{N}= #1 $}
\newcommand{\sub}[2]{#1_\text{#2}}
\newcommand{\partfrac}[2]{\frac{\partial #1}{\partial #2}}
\newcommand{\bsp}[1]{\begin{equation} \begin{split} #1 \end{split} \end{equation}}
\newcommand{\calF}{\mathcal{F}}
\newcommand{\calO}{\mathcal{O}}
\newcommand{\calM}{\mathcal{M}}
\newcommand{\calV}{\mathcal{V}}
\newcommand{\bbZ}{\mathbb{Z}}
\newcommand{\bbC}{\mathbb{C}}
\newcommand{\cK}{{\cal K}}

\newcommand{\Thq}{\Theta\left(\r-\r_q\right)}
\newcommand{\Dq}{\d\left(\r-\r_q\right)}
\newcommand{\kten}{\kappa^2_{\left(10\right)}}
\newcommand{\pbi}[1]{\imath^*\left(#1\right)}
\newcommand{\ho}{\hat{\omega}}
\newcommand{\tth}{\tilde{\th}}
\newcommand{\tf}{\tilde{\f}}
\newcommand{\tj}{\tilde{\j}}
\newcommand{\tw}{\tilde{\omega}}
\newcommand{\tz}{\tilde{z}}
\newcommand{\prj}[2]{(\partial_r{#1})(\partial_{\j}{#2})-(\partial_r{#2})(\partial_{\j}{#1})}
\def\atanh{{\rm arctanh}}
\def\sech{{\rm sech}}
\def\csch{{\rm csch}}
\allowdisplaybreaks[1]

\numberwithin{equation}{section}

\newcommand{\Tr}{\mbox{Tr}}    


%
\renewcommand{\theequation}{{\rm\thesection.\arabic{equation}}}
\begin{titlepage}
\rightline{MAD-TH-11-09}
\vspace{0.1in}

\begin{center}
\Large \bf  A tale of two cascades: Higgsing and Seiberg-duality cascades from type IIB string theory
\end{center}
\vskip 0.2truein
\begin{center}
Eduardo Conde$^{a,}$\footnote{eduardo@fpaxp1.usc.es}, 
J\'er\^ome Gaillard$^{b,}$\footnote{jgaillard@wisc.edu},
Carlos N\'u\~nez$^{c,}$\footnote{c.nunez@swansea.ac.uk},\\ ~ \\
Maurizio Piai$^{c,}$\footnote{m.piai@swansea.ac.uk} and
Alfonso V. Ramallo$^{a,}$\footnote{alfonso@fpaxp1.usc.es}
\vskip 0.2truein
 \vskip 4mm
 \it{$a$:
Departamento de  F\'\i sica de Part\'\i  culas\\ and\\ Instituto Galego de F\'\i sica de Altas
Enerx\'\i as
(IGFAE), \\ Universidade
de Santiago de
Compostela
\\E-15782, Santiago de Compostela, Spain}
\\
\vspace{0.2in}
{\it $b$: Physics Department.
University of Wisconsin-Madison
1150 University Avenue\\
Madison, WI 53706-1390. USA}

\vspace{0.2in}
{\it $c$: Department of Physics, Swansea University\\
 Singleton Park, Swansea SA2 8PP, United Kingdom.}
\vskip 5mm

\vspace{0.2in}
\end{center}
\vspace{0.2in}
\centerline{{\bf Abstract}}
We construct explicitly new solutions of type IIB supergravity 
with brane sources, the duals of which are ${\cal N}=1$ 
supersymmetric field theories exhibiting two very interesting phenomena.
The far UV  dynamics is controlled by a cascade of 
Seiberg dualities analogous to the Klebanov-Strassler 
backgrounds. At intermediate scales a cascade of 
Higgsing appears, in the sense that 
the gauge group undergoes a sequence of spontaneous 
symmetry breaking steps which reduces
its rank. Deep in the IR, the theory confines, 
and the gravity background has a non-singular end of space. 
We explain in detail how to generate such solutions, 
discuss some of the Physics associated with them
and briefly comment on the possible applications.

\smallskip
\end{titlepage}
\setcounter{footnote}{0}

\tableofcontents

\setcounter{footnote}{0}
\renewcommand{\theequation}{{\rm\thesection.\arabic{equation}}}
\section{Introduction }
The rigorous way to define an interacting field theory follows the 
 Wilsonian perspective~\cite{wilson}.
 One starts from a conformal theory, with a given basis of operators,
 and assumes that this provides a description of the dynamics of a fixed point
 which approximates the behavior at very short distances of the
 field theory of interest (in the case of asymptotically free 
 field theories, the conformal theory is actually free).
 One then adds deformations by turning on the couplings 
 of relevant operators of the conformal theory
(or vacuum expectation values for operators), which trigger
a renormalization group flow to low energies (longer distances).
The flow starts in the far ultraviolet from the original CFT (which is a fixed point of the RG flow)
and in the deep infrared can either end into another non-trivial
fixed point, or (in the case of confinement) into a trivial theory.
The RG flow itself, and the equations governing it,
encode all the most important dynamical properties of the field theory,
and studying it in detail provides a very clear 
strategy to characterize the field theory of interest.
However, these flows are typically difficult to study with standard
field-theoretic tools, in particular
because at some stage of the RG flow (if not all along the flow)
the field theory may be strongly coupled.

The AdS/CFT correspondence~\cite{hep-th/9711200}
conjectures the existence of a dual description 
to a conformal field theory in terms of a string-theory background 
with anti-de Sitter geometry, and its validity is supported by 
countless cross-checks.
It hence naturally offers a new opportunity to study non-trivial RG flows
by making use of  what today are called gauge-string dualities~\cite{hep-th/9802042},
thereby providing a very powerful analytical
 tool to deal with the study of RG flows in strongly coupled theories.
 Indeed,  many examples of flows like the one described above have been constructed.
 The basic idea behind such constructions is that one 
finds solutions to the string equations of motion which have a 
non-trivial dependence on a radial direction $\r$ in the geometry,
and interprets $\r$ as the renormalization scale in the RG flow
of some dual effective theory.
Provided the space asymptotes to
an anti-de Sitter geometry at large $\r$ (in the UV),
this procedure can be interpreted as providing the dual
description of the RG flow of a fundamental field theory.

A conceptually clean realization of these ideas is given
by the flow from $N=4$ Super-Yang-Mills to a theory with
minimal SUSY called $N=1^*$ Yang-Mills.
In the dual string description, the asymptotic space
is $AdS_5\times S^5$ (with a  flux for a RR five-form). This 
is deformed by the presence of non-trivial
NSNS $H_3$ and RR $F_3$ fields~\cite{hep-th/0003136},
and as a result the $\r$-dependence of the background is very non-trivial,
and reproduces the features expected from the RG flow of the dual field theory
(see also~\cite{PW} for the dual description of the flow resulting from
a very special deformation of ${\cal N}=4$).
This statement can be tested, because the preservation of some amount of supersymmetry
yields a degree of analytic control over the coupled non-linear differential
equations describing the flow; and hence certain calculations can
be performed on both sides of the duality, and compared.

Another non-trivial and successful application of these ideas has been developed by Klebanov
and collaborators, and relies instead on the $AdS_5\times T^{1,1}$ geometry, 
where $T^{1,1}$ is the base of the conifold. The dual conformal theory
is in this case  a minimally supersymmetric conformal field theory, described by a two-node
quiver with gauge groups of equal ranks and with a given number of bi-fundamental matter fields
\cite{hep-th/9807080}.
A deformation is introduced in the string background that
corresponds to an imbalance in the ranks of the gauge groups on the field-theory side of the correspondence.
 This yields an RG flow~\cite{hep-th/0002159} that leads to confining field theories, in either mesonic
or baryonic branches of the moduli space of the dual theory~\cite{KS,Dymarsky:2005xt},  
depending on the relations between the imbalance
and the original rank.
The downside of this construction, with respect to the previous case,
 is that the deformation does not switch itself off. 
 Once a non-trivial RG flow appears,
it does not reach a conformal
point at high energies. Instead, the flow follows closely a line of fixed points
described by  quiver field theories whose rank
grows without bound. Yet,  by calculating  correlators
of some key-operators~\cite{hep-th/0608209},
one can see that  the resulting system behaves in a way which resembles very closely
 a four-dimensional field theory.

This particular set-up ---
that we will refer to as the Klebanov-Strassler (KS) system ---
gave numerous insights into the strongly coupled
dynamics of four-dimensional field theories. These insights
could find applications in Particle Physics and Cosmology, aside from giving
non-trivial examples of  calculable strongly-interacting field theories.

The next stage in this descendent scale of 
`conceptually clean' set-ups is represented
by the so-called wrapped-brane models. 
Examples of these are the ones presented
in~\cite{hep-th/9803131,MN}, and various others. 
The field theory here is obtained by a twisted KK compactification of
a higher-dimensional field theory on a (curved) manifold.
A weak-coupling analysis of the 
dual theory shows that it contains an infinite tower of equidistant modes (see for example~\cite{AD}), 
revealing the existence of extra dimensions.
Supersymmetry and R-symmetry are (partially) broken and the RG flow of the four-dimensional
field theory never reaches a fixed point at high energies.
These models can be thought of as dual to effective theories which require 
a UV completion --- often non field-theoretic, after which a CFT is typically reached.
See~\cite{Bigazzi:2003ui}, for a detailed presentation
of each of the set-ups described above.
A very interesting fact is that the deep-IR behavior of backgrounds in this class
resembles very closely the backgrounds of models of the previous classes,
while the major differences arise only at large $\r$. 
Also, these models are comparatively simpler to construct and to study,
which means that they are very suitable for applications.

Another development that is relevant to this paper 
goes in the direction of constructing the duals of field theories 
with matter content in the fundamental representation of the gauge groups.
Adding $N_f$ fundamental fields in a  field theory with gauge group
$SU(N_c)$ is by now a well-studied problem in the context
of gauge/string duality. Reviews of different dynamical regimes, depending
on the ratio $\frac{N_f}{N_c}$, their string-theoretical description
 and applications can be found in
\cite{arXiv:0711.4467}-\cite{arXiv:1110.1744}.
It is in general very difficult, within this context, to 
construct backgrounds with the nice UV-asymptotic properties
of the dual to a fundamental field theory.

In this paper we will construct a set of solutions that fall
within the class of Klebanov-Strassler-like
backgrounds, in the sense that the string background 
is very closely related to the KS one far in the UV
and deep in the IR (i.e.~the backgrounds 
have the same good features as the KS ones at the extrema of the RG flow).
The novelty is given by the fact that the presence of $N_f$
sources, being $N_f\sim N_c$, modifies the ranks of the
gauge groups, and drastically changes
the RG flow at intermediate scales. 

We will also see that such backgrounds 
are non-trivially related, at least at the technical level,
 to the wrapped-brane models,
in the sense that they can be constructed by starting from one such model,
modified by the addition of $N_f$ sources, and by
applying an algebraic procedure that amounts to a partial
UV completion.  Notice however that the actual backgrounds of interest 
in this paper can also 
be constructed directly, without ever referring to the wrapped systems,
and that in this sense one might think of the reference to 
wrapped-brane systems as purely technical.
But it is suggestive, at least from the point of view of effective field theories, that 
finding these solutions is technically much easier 
by starting from the wrapped-brane systems.
\subsection{General idea of this paper}

Using a solution generating technique presented in
\cite{MM} and further developed in
\cite{Elander:2011mh}-\cite{rotations} (sometimes referred to as `rotation')
we construct a solution that generalizes
that of KS \cite{KS} and the so-called baryonic branch
of \cite{BGMPZ}
by the addition of source D5/D3-branes in the bulk that translate into 
a new
numerology for the field-theory quiver. The field theory, as we will see,
makes transitions between the numerology characterizing 
mesonic and baryonic branches and shows interesting
phenomena. From the view point of the string background,
it is completely smooth and  solves the equations of motion of
type IIB string theory coupled to the action of (smeared) sources. We
use techniques developed in previous
works to give a `profile' to the sources avoiding any singular behavior.

We analyze the field theory, whose dynamics includes
two cascades: one is the familiar cascade of Seiberg dualities
that is present in~\cite{KS}, while the other is 
associated with each crossing
of a source D3-brane in the radial direction (corresponding to
a  Higgs mechanism) as follows from the
ideas of \cite{hep-th/0101013}. We present various matchings between
field-theory expectations and calculations with
the string background. We also speculate about a possible
interesting application to the phenomena of
{\it Tumbling} in theories of Extended Technicolor.

In practical terms, what we are doing is the following.
We start from the wrapped-D5 system, for which it is comparatively simple to
find background solutions. We add flavor branes, using a profile 
for the sources such that the deep IR and far UV are virtually unaffected,
and solve the resulting background equations. 
We then apply the rotation procedure, in such a way as to cure the ills of the
wrapped system in the far UV, without affecting the deep IR.
The result is a very novel dynamical model, which yet preserves the good features of
older constructions.

Because we believe the results we obtain are very interesting,
in spite of the fact that at the technical level what we did
reduces to putting together a set of well-known ideas, 
the work is presented in full detail and a fair amount of background material
is also given, which makes the paper self-contained, if long. In a companion shorter paper \cite{skeleton}, we explore a particular realization of the ideas described above, in which the two cascades occur at different scales. 
The organization of the manuscript 
is the following: in Section \ref{section2}, we
present in detail the system of wrapped branes that is at the heart of our
treatment. We also review the application of the
solution generating technique of \cite{MM},
\cite{GMNP}. The techniques that allow to give
a profile to our sources are described in Section \ref{AEJ}.
Some new solutions
are displayed in that same section 
while some other new solutions to the type IIB equations
with sources are discussed in Section \ref{semianalyticdetails}. Field-theoretical quantities are analyzed in Section \ref{QFTQD} and some
applications are discussed in Section \ref{SECTIONAPP}.
We close the paper with conclusions in
Section \ref{sectionconclusions}. Generous appendices complement the presentation.
\section{Presentation of the SUSY system}\label{section2}
\setcounter{equation}{0}
In this section, we briefly  summarize some well-established 
aspects of two particular four-dimensional
supersymmetric field theories and their 
dual backgrounds. 
The two theories are  the Klebanov-Strassler \cite{KS},\cite{BGMPZ} 
QFT (referred below 
as `theory I') and the theory obtained when wrapping $N_c$ D5-branes on the 
two-cycle of the resolved conifold \cite{MN}, \cite{AD} 
(which we will refer to as `theory II'). We assume in the following 
that the reader is familiar with aspects of these two systems, and what follows is just a 
succinct account of their properties; some reviews are
\cite{Bigazzi:2003ui}.
The theory I
is a quiver with gauge group 
$SU(n+N_c)\times SU(n)$ and 
bi-fundamental matter multiplets $A_i, B_\alpha$ with $i,\a=1,2$.
The  global symmetries are\footnote{The R-symmetry is anomalous, breaking $U(1)_R\to Z_{2N_c}$.}
\beq
SU(2)_L\times SU(2)_R \times U(1)_B \times U(1)_R \,.
\label{globalsymm}\eeq
There is also a superpotential of the form $
W=\frac{1}{\mu}\epsilon_{ij}\epsilon_{\a\beta}tr[A_iB_\a A_j B_\beta].
$
The field theory is taken to 
be close to a strongly coupled fixed point at high energies. 
The dual description is given by 
the Klebanov-Strassler background \cite{KS} and its
generalizations \cite{BGMPZ}.
The field theory II 
has one gauge group $SU(N_c)$
and the global symmetries are 
those of eq.(\ref{globalsymm}) --- except for the baryonic
symmetry that is not present in this system.
These two theories, apparently so different, can be connected as discussed in
\cite{MM} and \cite{Elander:2011mh}  via Higgsing.
Indeed, giving a particular (classical) baryonic VEV to the fields 
$(A_i, B_\a)$ and expanding around it, the field content and degeneracies
of \cite{AD} are reproduced --- see \cite{MM}.
This weakly coupled field theory connection
has its counterpart in the type IIB solutions dual to each 
of the field theories.
Indeed, it is possible to connect the dual backgrounds 
of field theories I and II, 
using U-duality \cite{MM}. This connection was further studied in
\cite{Elander:2011mh}, \cite{GMNP}, \cite{rotations}.
In order to lay-out some background formalism and 
make manifest the connection between theories I and II
at strong coupling, we start from the type IIB configuration 
describing the strong dynamics of  the `field theory II' 
(a twisted compactification of D5-branes to four dimensions). It consists
of a metric, dilaton $\Phi$ and RR-three form $F_3$.
A quite generic background of this kind can be compactly 
written using the
$SU(2)$ left-invariant one-forms
\beq\lab{su2}
\tilde{\w}_1= \cos\psi\, d\tilde\theta+\sin\psi\sin\tilde\theta
\,d\tilde\varphi\,,\quad
\tilde{\w}_2=-\sin\psi\, d\tilde\theta+\cos\psi\sin\tilde\theta\,
d\tilde\varphi\,,\quad \tilde{\w}_3=d\psi+\cos\tilde\theta\, d\tilde\varphi\,,
\eeq
and the vielbeins
\beq\bal
&E^{x^i}= e^{\frac{\Phi}{4}} dx^i\,    ,\qquad
E^{\r}=  e^{\frac{\Phi}{4}+k}d\r\,  ,\qquad
E^{\theta}=  e^{\frac{\Phi}{4}+h}d\theta\,  ,\qquad
E^{\varphi}= e^{\frac{\Phi}{4}+h} \sin\theta\, d\varphi \,, \\
&E^{1}=  \frac{1}{2}e^{\frac{\Phi}{4}+g}(\tilde{\omega}_1 +a\, d\theta)\,  ,\;\;
E^{2}=\frac{1}{2}e^{\frac{\Phi}{4}+g}(\tilde{\omega}_2 
-a\,\sin\theta\, d\varphi)\,,\;\; E^{3}= \frac{1}{2}e^{\frac{\Phi}{4}+k}
(\tilde{\omega}_3 +\cos\theta\, d\varphi)\,.
\label{vielbeinbefore}
\eal\eeq
In terms of these, the background and the RR three-form, in Einstein frame, read
\beq\begin{aligned}
ds_E^2	&= \sum_{i=1}^{10} (E^{i})^2\,,    \label{f3old}\\
F_3			&= e^{-\frac{3}{4}\Phi}\Big(f_1 E^{123}+ f_2 E^{\theta\varphi 
3}
+ f_3(E^{\theta23}+ E^{\varphi 13})+ 
f_4(E^{\r 1\theta}+ E^{\r\varphi 2})   \Big)\,, 
\end{aligned}\eeq
where we defined
\beq\begin{split}
&E^{ijk..l}=E^{i}\wedge E^{j}\wedge E^{k}\wedge...\wedge E^{l}\, ,\\
&f_1=-2 N_c e^{-k-2g}\,,\qquad\qquad f_2= \frac{N_c}{2}e^{-k-2h}(a^2 -
2 a b +1 )\,,\\
&f_3= N_ce^{-k-h-g}(a-b)\,,\qquad f_4=\frac{N_c}{2}e^{-k-h-g}b' \,.
\end{split}\eeq
The system has a radial coordinate $\r$, on which $(a,b,\Phi,g,h,k)$ depend,
and we have set $\alpha^{\prime}g_s=1$.
The full background is then determined by 
solving the equations of motion for the 
functions $(a,b,\Phi,g,h,k)$. 
A system of BPS equations is derived using 
this ansatz (see appendix of
reference \cite{Casero:2006pt}). These non-linear and coupled 
first-order equations
can be arranged in a convenient form, by rewriting the
functions of the background in terms of a new basis
of functions that decouples the equations (as 
explained in
\cite{HoyosBadajoz:2008fw}-\cite{Casero:2007jj}). 
We quote this change of basis
in our Appendix \ref{backgroundfunctionsappendix}.
Using these new variables, one can solve for all of them
except for a function $P(\r)$, that is determined by a second-order equation,
see \cite{HoyosBadajoz:2008fw} 
and our Appendix \ref{backgroundfunctionsappendix} for details.
The second-order equation  reads
\begin{align}
& P'' + P'\Big(\frac{P'+Q'}{P-Q} +\frac{P'-Q'}{P+Q} - 4 
\coth(2\rho)
\Big)=0\,, \nonumber\\
& Q(\r)= N_c\,(2\r \coth(2\r)-1)\,.
\label{master}
\end{align}
We will refer to eq.(\ref{master}) 
as the {\it master equation}: this is the only 
equation that needs solving in order to 
generate the large classes of solutions of type IIB dual to
`field theory II' 
in different
vacua, deformations of the Lagrangian, VEV's, etc.
\subsection{Aspects of the SUSY solutions}
Let us start by describing  some known 
solutions of the master equation (\ref{master}).
As a first example, the simple solution $P=2N_c\, \r$ 
gives the background of \cite{MN},
\cite{Chamseddine:1997nm}. This solution and those with the same large-$\r$
asymptotics will not be the main focus 
of this paper.
Other interesting solutions can be found semi-analytically 
(with asymptotic series expansions
and numerical integration). Let us discuss 
the large and small-$\r$
expansions of the function $P(\r)$ that 
will be of interest in this work. 
For large values of the radial coordinate (describing 
the UV of the field theory II), the solution  
has an expansion  of the form,
\bea
&P&=e^{4\rho/3}\Big[ c_+  
+\frac{e^{-8\r/3} N_c^2}{c_+}\left(
4\r^2 - 4\r +\frac{13}{4} \right)+ e^{-4\r}\left(
c_- -\frac{8c_+}{3}\r \right)+\nonumber\\
& & + \frac{N_c^4 e^{-16\r/3}}{c_+^3}
\left(\frac{18567}{512}+\frac{2781}{32}\r +\frac{27}{4}\r^2 +36\r^3\right)  + \co(e^{-20\rho/3})
\Big]\,.
\label{UV-II-N}
\eea
Notice that this expansion 
involves two integration constants, $c_+>0$ and $c_-$. 
The background functions
at large $\r$ are written for reference
in Appendix \ref{backgroundfunctionsappendix}.
Regarding the IR expansion, 
we look for solutions with $P\to 0$ 
as $\r\to 0$, 
in which case we find
\beq
P= h_1 \r+ \frac{4 h_1}{15}\left(1-\frac{4 N_c^2}{h_1^2}\right)\r^3
+\frac{16 h_1}{525}\left(1-\frac{4N_c^2}{3h_1^2}-
\frac{32N_c^4}{3h_1^4}\right)\r^5+\co(\r^7)\,,
\label{P-IR}\eeq
where $h_1>2N_c$ is again an arbitrary constant; 
there is another integration constant, $P(0)$,
taken to zero here, to avoid singularities. In order to match the UV expansion, it amounts to taking as well $c_-=0$ in \eqref{UV-II-N}.
This gives background functions
that are quoted in Appendix \ref{backgroundfunctionsappendix}.
Of course, there is a smooth numerical interpolation between both 
expansions. As one can imagine, there is then only one independent 
parameter; given a value for one of the constants $\{c_+, h_1\}$, 
the requirement that the solution matches \emph{both} 
expansions is sufficient to determine the value of the other.
The small-$\r$ behavior of the expansion above 
is that
of the exact solution $P=2N_c \r$ (but notice that in eq.(\ref{P-IR})
the constant $h_1>2N_c$). In contrast,
the large-$\r$ expansion $P\sim e^{4\r/3}$ greatly differs from 
the linear behavior.
Let us focus our attention on the dilaton field in backgrounds like those in 
eq.(\ref{f3old}). For small and large 
values of the radial coordinate, we have
\bea
& & e^{4\Phi}|_{\r\to 0}=   
e^{4\Phi(0)}\Big[ 1+\frac{64  N_c^2 \rho ^2}{9 h_1^2}
+\frac{128 N_c^2 \left(-15 h_1^2+124 N_c^2\right) \rho ^4}{405 h_1^4}+
\co(\r^6)\Big]\,,\nonumber\\ 
& & e^{4\Phi}|_{\r\to \infty}=
e^{  4 \Phi(\infty)} \Big[1+\frac{3N_c^2 e^{-8\r/3}}{4c_+^2}(1-8\r)+ \co(e^{-16\r/3})
\Big],
\label{dilatonzzz}
\eea
and there is a relation between $\Phi(0)$ and $\Phi(\infty)$. Notice that the dilaton (that is related to the warp factor)
asymptotes to a constant at large values of the radial coordinate\footnote{Some readers may find this reminiscent of
what happens when `keeping the 1' in the D3-brane warp factor.}. 
One can see \cite{Elander:2011mh}
that this type of solutions corresponds to
the addition of an irrelevant operator in the Lagrangian of the theory
of the type II field theory, needing a UV completion.
\subsubsection{A  nice way of writing solutions}\label{exactformal}
We summarize here a curious way of finding
an exact and analytic solution to the master equation (\ref{master}),
in terms of an infinite series of non-explicit integrals. The usefulness of 
this formal solution will become
clear in the following sections.
We  rewrite the master eq.(\ref{master}) as
\beq
\partial_\r\Big( \frac{(P^2-Q^2)}{\sinh^2 (2\r)} P' \Big) 
+\frac{4}{\sinh^2 (2\r)}P'QQ'=0\,.
\eeq
We integrate this expression twice, taking already into account the asymptotics described above, to get
\beq
P^3 - 3 P Q^2 + 6 \!\int_0^\r d\tilde\r \, PQQ'-12\! \int_0^\r
d\tilde{\r}\,\sinh^2(2\tilde{\r})\int^{\infty}_{\tilde{\r}}d\hat{\r}\,
\frac{P' QQ'}{\sinh^2 (2\hat\r)}=4\l^3\epsilon^4\left( \int_{0}^{\r}d\tilde{\r} \sinh^2(2\tilde\r)\!\right),
\label{P_integrated}
\eeq
where we have defined for convenience the constant $\l=2^{\frac{2}{3}}c_+ \epsilon^{-\frac{4}{3}}$, where $\epsilon$ will be later identified with the deformation parameter of the deformed conifold\footnote{The deformation parameter $\epsilon$ of the conifold is such that $\epsilon^{4/3}$ has dimension of length squared, the same as $c_+$ (restoring units of $g_s \a'$). Thus $\l$ is a dimensionless parameter.}.
We propose a solution in an inverse series expansion in the dimensionless constant
$\lambda$:
\beq
P= \lambda P_1 + P_0 + 
\frac{P_{-1}}{\lambda}+
\frac{P_{-2}}{\lambda^2}+
\frac{P_{-3}}{\lambda^3}+\ldots\,\,.
\eeq
Plugging this expansion in (\ref{P_integrated}) and 
matching the different powers of $\lambda$, 
we get
\bea
& & P_1= \left(4\epsilon^4  \int_{0}^{\r}\,d\tilde{\r} \sinh^2(2\tilde\r)\right)^{1/3},\nonumber\\\label{lambdaseries}
& & P_0= P_{-2}=\ldots=P_{-2k}=0\,,\\
& & P_{-1}= -\frac{1}{P_1^2}\Big(-
P_1Q^2+2\int_0^\r d\tilde\r \, P_1QQ'  -4 \int_0^{\r} d\tilde\r\, \sinh^2(2\tilde\r)\int_{\tilde\r}^{\infty} 
d\hat{\r}\,\frac{P_1'QQ'}{\sinh^2(2\hat\r)}\Big)\,.\nonumber
\eea
A recurrence relation for $P_{-(2k+1)}$ can be found
in eq.(B.7) of \cite{HoyosBadajoz:2008fw}. 
This series converges rapidly to the numerical solutions of eq.(\ref{master}) and 
is
a very good  approximation for large 
values of the radial coordinate.

An interesting limit to take  is $\lambda \to \infty$. We will 
explore this in the following sections.
\subsection{Connecting to Theory I}
As explained in \cite{Elander:2011mh}, the solution  
in eqs.(\ref{UV-II-N})-(\ref{P-IR}), supplemented with a 
suitable numerical interpolation,
describes the strong dynamics  
of the dual field theory II in the presence of a dimension-eight 
operator inserted in the Lagrangian (which ultimately couples 
the field theory to gravity).
This calls for a completion in the context of field theory which
is achieved with  the U-duality of \cite{MM} (a solution generating technique
sometimes referred to as `rotation').
After the U-duality described in \cite{MM} is applied, we
define the new vielbein 
(which we use in the following),
\begin{align}
	e^{x^i}			&= e^{\frac{\Phi}{4}}\hat{h}^{-\frac{1}{4}} dx^i    
\,,\;\;\;
	e^{\r}	 		=  e^{\frac{\Phi}{4}+k} \hat{h}^{\frac{1}{4}}d\r  
\,,\;\;\;
	e^{\theta}	=  e^{\frac{\Phi}{4}+h} \hat{h}^{\frac{1}{4}}    d\theta  
\,,\;\;\;
	 e^{\varphi}= e^{\frac{\Phi}{4}+h} \hat{h}^{\frac{1}{4}} 	\sin\theta \,
d\varphi\,    ,\nonumber\\
	e^{1}				&=  \frac{1}{2}e^{\frac{\Phi}{4}+g} 
\hat{h}^{\frac{1}{4}}	(\tilde{\omega}_1 +a\, d\theta)\,  ,\qquad\qquad
	e^{2}				=		\frac{1}{2}e^{\frac{\Phi}{4}+g} 
\hat{h}^{\frac{1}{4}}	(\tilde{\omega}_2 	-a\,\sin\theta\, d\varphi)   
\,,\nonumber\\ 
	e^{3}				&= \frac{1}{2}e^{\frac{\Phi}{4}+k} 
\hat{h}^{\frac{1}{4}}	(\tilde{\omega}_3 +\cos\theta\, d\varphi)\,.
	\label{vielbeinafter}
\end{align}
The dilaton and RR three-form are
invariant under this operation.
The newly generated  metric, RR and NSNS fields are
\bea
& & ds_E^2= \sum_{i=1}^{10} (e^{i})^2\,,\nonumber\\
& & F_3= \frac{e^{-\frac{3}{4}\Phi}}{\hat{h}^{3/4}}
\Big[f_1 e^{123}+ f_2 e^{\theta\varphi 3}
+ f_3(e^{\theta23}+ e^{\varphi 13})+ 
f_4(e^{\r 1\theta}+ e^{\r\varphi 2})   \Big]\,,\nonumber\\
& & B_2= \k\, \frac{e^{\frac{3}{2}\Phi}}{\hat{h}^{1/2}}\Big[e^{\r 3}-\cos\alpha
(e^{\theta\varphi}+ e^{12})-\sin\alpha(e^{\theta2}+ e^{\varphi 1})   
\Big]\,,\nonumber\\
& & H_3=-\k\, \frac{e^{\frac{5}{4}\Phi}}{\hat{h}^{3/4}}
\Big[-f_1 e^{\theta\varphi \r} - f_2 e^{\r 12}
- f_3(e^{\theta 2\r}+ e^{\varphi 1\r})+
f_4(e^{ 1\theta 3}+ e^{\varphi 2 3})   \Big]\,,\nonumber\\
& & C_4= -\k\,\frac{e^{2\Phi}}{\hat{h}} 
dx^0\wedge dx^1 \wedge dx^2\wedge dx^3\,,  \nonumber\\
& & F_5= \k\, e^{-\frac{5}{4}\Phi -k}\hat{h}^{\frac{3}{4}}
\partial_\r \left(\frac{e^{2\Phi}}{\hat{h}}\right)
\Big[e^{\theta\varphi 123 }- e^{x^0 x^1 x^2 x^3 \r}   \Big]\,.
\label{configurationfinal}
\eea
We have defined
\beq
\cos\alpha= \frac{\cosh(2\r)-a}{\sinh(2\r)}\,,\qquad \sin\alpha= -\frac{2e^{h-g}}{\sinh(2\r)}\,,\qquad\qquad\hat{h}=1-\k^2 e^{2\Phi}\,,
\label{cosinh}
\eeq
where $\k$ is a constant that we will choose to be $\k= e^{-\Phi(\infty)}$,
requiring the dilaton to be 
bounded at large distances\footnote{This is the restriction on the type 
of solutions considered that rules out the exact solution $P=2N_c\r$.}.
The rationale for this choice is to obtain 
a dual QFT that is UV-complete, see \cite{Elander:2011mh}
for
explanations. 
The background in eq.(\ref{configurationfinal}) has the same 
form as the one describing the baryonic branch
of the KS theory \cite{BGMPZ}. Indeed, one can check that the BPS 
equations written in \cite{BGMPZ} are exactly equivalent to our 
eq.(\ref{master}). 
Another way of understanding the connection was explained in
\cite{GMNP}: what relates the backgrounds in eqs.(\ref{f3old})
and (\ref{configurationfinal}) is a rescaling of the almost K\"ahler 
and complex structure forms describing 
the six-dimensional internal space. 

A closer inspection strongly suggests that 
the formal solution written in Section \ref{exactformal}
is the solution of \cite{BGMPZ}.
Let us study the limit $\lambda\to \infty$ 
of the special solution presented in Section \ref{exactformal}. 
In that case the solution reduces to 
\beq
P\sim \l P_1= \left(4\epsilon^4 \int \sinh^2(2\r)\right)^{1/3}= \l\, 2^{-\frac{1}{3}}\epsilon^{\frac{4}{3}}(\sinh(4\r)-
4\r)^{1/3},
\label{P1}
\eeq
For the different background functions the limit results in 
\beq\begin{aligned}
& e^{2h}= \l \frac{\tanh(2\r)}{4}P_1+\cO(\l^0)\,,
&& \frac{e^{2g}}{4}= \l \frac{\coth(2\r) P_1 }{4}+\cO(\l^0)\,,
\\
& \frac{e^{2k}}{4}= \l \frac{P_1'}{8}+\cO(\l^{-1})\,,
&& a= \frac{1}{\cosh (2\r)}+\cO(\l^{-1})\,,\\
& b= \frac{2\r}{\sinh(2\r)}\,,
& & e^{4\Phi-4\Phi_0}= \frac{2\sinh^2(2\r)}{\l^3 P_1^2 P_1'}+\cO(\l^{-5})\,,\\
\end{aligned}\label{functionslambdainfinity}\eeq
\beq
\hat{h}= 1- \frac{\k^2 e^{2\Phi_0}}{\l^{3/2}}\sqrt{\frac{2\sinh^2(2\r)}{P_1^2P_1'}} 
+ \frac{\k^2 e^{2\Phi_0}}{\l^{7/2}}\frac{\sinh(2\r)(2P_1P_1'P_{-1}+P_1^2P'_{-1}-P_1'Q^2)}{\sqrt{2P_1'}P_1^3}+\co(\l^{-11/2})\,. \nonumber
\eeq
The choice of $\k=e^{-\Phi(\infty)}$ implies that
$3\k^4= 2\epsilon^4\l^{3}e^{-4\Phi_0}$, and the logic for this choice is to get a 
warp factor that vanishes at large $\r$. Indeed, plugging this value in eq.\eqref{functionslambdainfinity}, and using the explicit expression of $P_1$ given in eq.\eqref{P1}, we see that the two first terms in the expansion for $\hat{h}$ cancel out and
\beq
\hat{h}= \frac{1}{\l^{2}}\frac{\epsilon^2\sinh(2\r)(2P_1P_1'P_{-1}+P_1^2P'_{-1}-P_1'Q^2)}{P_1^3\sqrt{3P_1'^3}}+\co(\l^{-4})=\frac{\hat{h}_2}{\l^2}+\co(\l^{-4})\,.
\label{hhat}
\eeq
This in turn
implies that we are switching off a dimension eight operator
in the dual QFT, obtaining a proper UV-complete field theory. A 
detailed discussion of this choice, showing that the decoupled operator 
has dimension eight,
is in \cite{Elander:2011mh}. 
Regarding $\hat{h}_2$
we explicitly have,
\beq
{\hat{h}_2}= \frac{2^{5/3}}{\epsilon^{8/3}} N_c^2 \int_{\r}^{\infty} dx
\frac{(\sinh(4x)-4x)(2x \coth(2x)-1)}
{\sinh^2(2x)( \sinh(4x) -4x)^{2/3}}\,,
\label{h2}
\eeq
which is the Klebanov-Strassler \cite{KS} warp factor $\hat{h}_2=\hat{h}_{KS}$.
Finally, rescaling the Minkowski coordinates
$x^i\to x^i \l^{-1/2}$ the metric is independent of the parameter
$\l$, and  using that
\beq
\frac{P_1'}{P_1}=\frac{8 \sinh^2(2\r)}{3(\sinh(4\r)-4\r)}\,,
\eeq
the internal space metric is the deformed conifold.
This is {\it precisely} the Klebanov-Strassler solution \cite{KS} that 
is obtained as  the limit $\l\to\infty$ of the formal solution in
Section \ref{exactformal} (the latter is the full baryonic branch 
solution \cite{BGMPZ}).
\subsection{Adding sources}\label{additionsourceszzz}
We consider now the addition of sources in these backgrounds
and the effect of the rotation on them.
The logic 
followed is the one described in \cite{Casero:2006pt}, \cite{arXiv:1002.1088}.
We  add a bunch of $N_f$ D5-branes in backgrounds of the form 
given in eq.(\ref{f3old}). 
These D5-branes are added as explained in \cite{Casero:2006pt}.
They  are in principle localized in the compact space
described by the angles $[\theta,\varphi, \tilde{\theta},\tilde{\varphi}]$.
One can run the machinery described in \cite{GMNP} and  
\cite{Martucci:2005ht}, 
to write  
a BPS system of nonlinear coupled partial 
differential equations (with derivatives on the radial coordinate and the
angles where the sources are localized). The 
difficulty of the problem
prompted the authors of 
\cite{Bigazzi:2005md},
\cite{Casero:2006pt}, to {\it smear} the sources\footnote{Equivalently, 
either consider the Fourier zero mode of the solution
in the compact internal space, or consider a situation 
where the global symmetries of the field theory without 
flavors/sources --- eq.(\ref{globalsymm}) --- are 
respected after the addition of sources. See \cite{arXiv:1002.1088} 
for discussions on different aspects of the smearing.}.
In this smeared, $SU(2)_L\times SU(2)_R$ invariant situation
one can run a very similar program to the one described above:
generalize the change of variables discussed in Appendix
\ref{backgroundfunctionsappendix} and write a 
new master equation for the
function $P(\r)$,
\bea
& & P''+(P'+N_f )\Big[\frac{P'+Q'+2 N_f }{P-Q}
+ \frac{P'-Q'+2 N_f }{P+Q}
- 4 \coth(2\r)   \Big]=0\,,\nonumber\\
& & Q(\r)= \frac{2N_c-N_f}{2}(2\r \coth(2\r)-1)\,.
\label{mastereq}
\eea
See the paper \cite{HoyosBadajoz:2008fw} for technical details.
In this case, solutions are known either 
semi-analytically (as series expansions plus a smooth 
numerical interpolation) and also generalizing
the formalism of Section \ref{exactformal}.
A classification of the solutions according to the asymptotic behavior of 
$P(\r)$, 
together
with expansions of all other 
background functions in the presence of sources
was given in  \cite{HoyosBadajoz:2008fw}. We will not 
quote the results here, but refer the reader to
\cite{HoyosBadajoz:2008fw} and \cite{GMNP} for details.
A feature of all these solutions is the presence of singularities for
small $\r$. In other words, one finds a divergent Ricci scalar as 
$R\sim \frac{N_f}{\r^2}$.
A physical reason of the existence of the singularity 
(namely a very large density of `crossing' source-branes 
at the point $\r=0$) was given in
\cite{arXiv:1002.1088}, among other papers.
More in detail, the solutions in the presence of the D5 sources are
formally like those in eq.(\ref{f3old}), with the difference that the functions
$f_1, f_2,f_3, f_4$ change to reflect the presence of sources ($dF_3$ is 
nonzero).
Indeed, we now have,
\bea
& & f_1=-2 N_c \,e^{-k-2g}\,,\;\;\; \qquad f_2= \frac{N_c}{2}e^{-k-2h}(a^2 -
2 a b +1 -\frac{N_f}{N_c})\,,\nonumber\\
& & f_3= N_c e^{-k-h-g}(a-b)\,,\;\;\; \qquad f_4=\frac{N_c}{2}e^{-k-h-g}\,b'\, .
\label{malala}\eea
The other functions in the background also reflect the presence of the $N_f$
sources. 
After the rotation procedure is applied we will have a background
that formally is the one in eq.(\ref{configurationfinal}), 
but with the functions
$\Phi,h,g,k,a,b$ that now have 
contributions coming from the sources.
There are two interesting points to discuss. First, as explained in 
detail in \cite{GMNP},  the presence of the
NSNS
$B_2$ field will induce D3-brane charge on the D5 sources after the rotation. 
The second point is that due to the presence of D5 sources
the function $P(\r)$ gets a contribution at large values of the 
radial coordinate, see \cite{HoyosBadajoz:2008fw} 
and compare with eq.(\ref{UV-II-N}),
\beq
P(\r)= e^{4\r/3}\Big(c_+ +\frac{9N_f}{8}e^{-4\r/3}+\cO(e^{-8\r/3})    \Big)\,.
\eeq
Carefully following the details --- see \cite{GMNP} --- one finds 
that this new term impacts the large-radius asymptotics of
the warp factor for the rotated background 
as
\beq
\hat{h}= N_f e^{-4\r/3} + \frac{(2N_c-N_f)^2}{2c_+} \r\, e^{-8\r/3}+\cO(e^{-8\r/3})\,.
\label{xxyyzz}
\eeq
Then, asymptotically, the 
cascading behavior of the KS system, represented above by the
$e^{-8\r/3}$ term present also in $\hat{h}_2$ of eq.(\ref{h2}), 
is overcome by the presence 
of the D3 and D5 sources. This deviates the system from the `near $AdS_5$' UV.
\subsection{About the dual field theory}
We comment briefly on the field theory dual to the (singular) geometry
after the rotation with sources. This was proposed in \cite{GMNP} to 
be the quiver
\beq
SU(n+ N_c+n_f)\times SU(n+n_f+\frac{N_f}{2})\,,
\eeq
($n_f$ is the number of D3-branes induced by the sources, more precisions to be given
in the following sections) with 
superpotential $W\sim ABAB$
for the fields $A_i, B_\alpha$ transforming as bi-fundamentals of the gauge groups,
in the same fashion as indicated around eq.(\ref{globalsymm}).
It was also proposed that the field theory
was on a {\it mesonic branch} and that as discussed in \cite{Dymarsky:2005xt}
the baryonic symmetry $U(1)_B$ was unbroken. Computations 
of the $c$-function, beta functions and anomalies supported this proposal.
The presence of the small-$\r$ singular behavior of the background (inherited
from the singularity in the unrotated background), 
implies that we should not trust
the IR dynamics as read from the geometry. At best, the solution should be
thought of as a generalization of the Klebanov-Tseytlin geometry
\cite{hep-th/0002159} in the presence of extra matter/sources. 
This point was made concrete by calculating Wilson loops that displayed a 
non-physical behavior for large separations \cite{Bennett:2011xd}.

It was clearly stated in \cite{GMNP} that a resolution of the singularity
was very desirable as it would allow to trust the low energy dynamics
of the field theory read from the type IIB background with sources.
Here, we also point to the need of having a better understanding of the
high-energy (and strongly coupled) dynamics. As pointed out 
above, this dynamics seems to be driven by higher-dimensional operators reflected in
the term $N_f\, e^{-4\r/3}$ in eq.(\ref{xxyyzz}).
We will comment and resolve  
both issues in the following sections.
\section{Solving the problems: sources with a profile}\label{AEJ}
\begin{figure}[h]
\begin{center}
\includegraphics[width=0.5\textwidth]{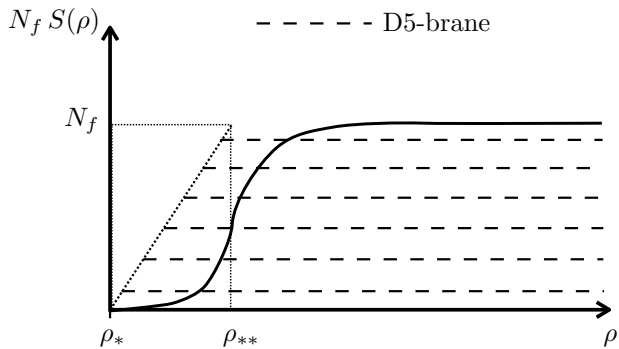}
\caption{We plot an example of a possible D5-brane configuration: a total of $N_f$ D5s extend in the radial coordinate from some minimum $\rho$, that can lie between $\rho_*$ and $\rho_{**}$, to infinity. We chose a distribution of minimal radial distances of the D5s that is homogeneous in $\rho$, as indicated by the dotted oblique line. This configuration of D5s gives rise to the plotted $S(\rho)$, which is asymptotically constant ($S(\rho)\to1$).}
\label{fig:D5s}
\end{center}
\end{figure}

\setcounter{equation}{0}
Let us briefly remind the reader about the recent results of 
\cite{Conde:2011rg}. The paper is written in the context of
the backgrounds of eq.(\ref{f3old}) for $(N_c, N_f)$ D5 color
and flavor branes. The authors found that a way of getting rid 
of the small-radius singularity was to avoid the `very high density'
superposition of sources (as described above and also pointed out in
\cite{arXiv:1002.1088}, \cite{Benini:2006hh}). 
For this, the authors of \cite{Conde:2011rg}
developed a formal way of distributing the sources such that
not all the $N_f$ branes reach the point $\r=0$.
They constructed a distribution of sources (flavors) with a particular profile
that vanishes near the origin and 
stabilizes to a constant at large radius as indicated in the Figure \ref{fig:D5s}.

The distribution $S(\r)$ is obtained rigorously solving the 
kappa symmetry equations for the branes
and distributing them following a well-defined procedure
\cite{Conde:2011rg}, \cite{Barranco:2011vt}. 
This distribution can be interpreted in
two ways: 
the first,  as the addition of radius-energy-dependent  
masses for the flavors in 
the dual field theory --- the profile of masses is somewhat related to 
$N_f\,S(\r)$. The second way of interpreting the distribution of sources,
 uses that the field theory II couples to flavors
according to the 
superpotential (see \cite{Casero:2006pt},
\cite{Casero:2007jj})
\beq
W\sim \tilde{Q}^\dagger \hat{\Phi}_k Q\,,
\eeq
where ($\tilde{Q}$) $Q$ are the (anti) quark superfields 
introduced by the sources
and  
$\hat{\Phi}_k$ is a generic massive chiral 
multiplet in the field theory II. The profile is understood
as an energy-dependent VEV for 
the field $\hat{\Phi}_k$, that effectively generates
an energy-dependent mass term for the flavor multiplets. 
While the outcome in both cases
is an energy-dependent
 `distribution of mass', the second interpretation 
implements it without 
breaking the R-symmetry, hence making it our choice.
Let us describe how this distribution of sources 
affects the string background.

Indeed, one writes a
background like that in eq.(\ref{f3old}), where the functions
$f_1,\ldots,f_4$ now reflect the presence of smeared sources 
with a profile $S(\r)$
\footnote{Compare this to eq.(\ref{malala}) 
to notice that the effect of the
profile for the sources is {\it not} just the change $N_f\to N_f S(\r)$.
See \cite{Conde:2011rg} for details.}:
\beq\bal
& f_1=-2 N_c e^{-k-2g}\,, &f_2= \frac{N_c}{2}e^{-k-2h}\left( a^2 - 2 a b +1 -\frac{N_f}{N_c}S(\r)\right)\,,\\
& f_3= N_c\, e^{-k-h-g}(a-b)\,,& f_4=\frac{N_c}{2}e^{-k-h-g}\left( b'+\frac{N_f S'(\r)}{2N_c \cosh(2\r)}\right)\, .
\eal\label{ffs}
\eeq
Following 
\cite{Conde:2011rg}, 
one can write the BPS equations 
for the system with
the sources added with a profile. Here again, the formalism of 
\cite{HoyosBadajoz:2008fw} can be generalized, in this case for a distribution
$N_f(\r)=N_f S(\r)$.
Indeed, all functions can be immediately obtained once we know
the functions $P,Q$. Following the treatment in 
\cite{Conde:2011rg} we learn that
\beq
Q\,=\,\coth(2\r)\,\,\Big[\,\int_0^\r dx \,
\frac{2N_c\,-\,N_f\,S(x)}{\coth^2(2x)}\,\Big]\,\,,
\label{Q-integral}
\eeq
where an integration constant has been set to zero to avoid IR singularities. 
Moreover, as shown in \cite{Conde:2011rg}, 
one can find a master equation,
\begin{equation}
	P'' + N_f\,S' + (\,P' + N_f\,S\,) \left( 
\frac{P' - Q' + 2 N_f\,S}{P + Q} + 
\frac{P' + Q' + 2 N_f\,S}{P - Q} - 4 \coth (2\r) \right) = 0\,\,,
\label{master-eq}
\end{equation}
which reduces to eq.(\ref{mastereq}) 
when $S(\r)=1$  and 
to eq.(\ref{master}) for $N_f=S(\r)=0$ .
 
\begin{figure}[h]
\begin{center}
\includegraphics[width=0.5\textwidth]{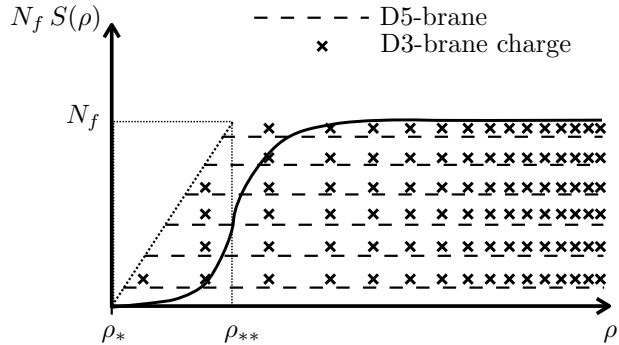}
\caption{After the rotation, D3-brane charge is induced on the D5s. 
As argued in Section \ref{twokinds}, this charge grows exponentially 
with the radial distance. An asymptotically constant 
profile $S(\rho)$ will generate an infinite amount D3-brane charge.}
\label{fig:D5sD3s}
\end{center}
\end{figure}

Once the function $S(\r)$ is known 
(from a microscopic  description of  the smearing of sources), 
one can  get $Q(\r)$ from (\ref{Q-integral}) and
solve the second-order master equation (\ref{master-eq})
for $P(\r)$.
The rest of the functions have expressions that generalize those in our
Appendix \ref{backgroundfunctionsappendix}, as
 we read from
\cite{Conde:2011rg},
\beq\bal
&e^{2h}=\frac{1}{4}\,\frac{P^2-Q^2}{P\coth(2\r)-Q}\,,&e^{2g}=P\,\coth(2\r)-Q\,,\\
&e^{2k}\,=\,\frac{P'+N_f\,S(\r)}{2}\,,&a=\frac{P}{P\cosh(2\r)-Q\sinh(2\r)}\,.
\eal\label{h-k-P-Q}
\eeq
The dilaton is,
\beq
e^{4\Phi-4\Phi_0}= \frac{2\sinh(2\r)^2}{(P^2-Q^2)(P'+N_f S)}\,.
\eeq
and the function $b(\r)$ has the form
\beq
b(\r)= \frac{2\r}{\sinh(2\r)}-\frac{N_f}{2N_c}\Big[\frac{S(\r)}{\cosh(2\r)}
+\frac{2}{\sinh(2\r)}\int_0^{\r}dx\,\tanh(2x)^2 S(x)   \Big].
\label{bderho}\eeq
The problem just boils down to finding $P(\r)$, given $S(\r)$.
We follow the treatment in 
\cite{Conde:2011rg} and 
\cite{Barranco:2011vt} --- where quite generic distributions
$S(\r)$ were constructed --- to get particular 
expressions for the function $S(\r)$
that make the expression of $Q(\r)$ analytic and the 
resolution of the master equation simple.
Performing the rotation is immediate,  and we will  quote below
the expression for the full rotated background. 
As we discussed, the rotation induced D3-brane charges
on the D5 sources. This is represented in Figure \ref{fig:D5sD3s}.

\subsection{The two kinds of solutions}\label{twokinds}
Given a distribution function for the D5 sources $S(\r)$, 
we can actually construct two kinds of solutions: those that are 
{\it exact and analytic} and those that are {\it semi-analytic} 
(in the sense of 
knowing either their asymptotic power series expansions or an exact 
expression in terms
of an infinite number of integrals that cannot be evaluated exactly).
\begin{itemize}
\item{Exact solutions: 
they are obtained in the double-limit 
$\l\to\infty$ and $N_f\to 0$, keeping $\lambda N_f$ fixed. We then define $\nu =2^{-2/3} \epsilon^{4/3} \lambda N_f= c_+ N_f$, where $c_+$ was the dimensionful constant in equation \eqref{UV-II-N}. The solutions are very similar to the 
KS background in the sense that 
functions are written up to an integral, the 
dilaton is constant and $F_3={*}_6 H_3$. 
In these solutions there are  D3 sources, but no D5 sources. 
}
\item{Semi-analytic solutions:
the parameters $\l, N_f$ 
in the expansion
are both fixed and nonzero. These solutions 
are  written 
in terms of an infinite sum of integrals. 
The treatment follows what is written in 
Appendix \ref{appendixdetailedderivation}. For all practical purposes, 
this is equivalent to writing asymptotic series for large and small values
 of $\r$ and a smooth numerical interpolation. In these solutions
there are D3 and D5 sources.
}
\end{itemize}
We will give details on the semi-analytic solutions in Section 
\ref{semianalyticdetails}.
In the case of exact solutions, the configuration in string frame reads 
(details of the derivation are given in Appendix 
\ref{appendixdetailedderivation})
\beq\begin{aligned}
ds^2&= \hat{h}^{-1/2}dx_{1,3}^2 + \hat{h}^{1/2}ds_6^2\,,\\
ds_6^2&=P_1 \Big(\frac{P_1'}{2P_1}
(d\r^2+\frac{1}{4}(\tilde{\omega}_3 +\cos\theta d\varphi)^2) + 
\frac{\tanh(2\r)}{4}(d\theta^2+\sin^2\theta d\varphi^2)\,+\\
&\quad+\frac{\coth(2\r)}{4}\left((\tilde{\omega}_1+a d\theta)^2+(\tilde{\omega}_2-
a \sin\theta d\varphi)^2\right)   \Big)\,,\\
B_2&=- \frac{N_c}{4} \frac{2\r \coth(2\r) -1}{\sinh(2\r)}\left(\cosh(2\r) \left(\sin \theta\, d\theta \wg d\varphi - \tilde{\omega}_1 \wg \tilde{\omega}_2\right)-\sin \theta\, d\varphi \wg \tilde \w_1 - d \theta \wg \tilde\omega_2 \right)\,,\\
H_3& =dB_2,\qquad F_3={*}_6 H_3\,,\\
F_5&=-(1+*)\partial_\r \hat{h}^{-1} dx^0\wedge dx^1\wedge dx^2\wedge dx^3 \
\wedge d\r\,,\\
\hat{h}&= \hat{h}_{2} + \frac{4}{\epsilon^{8/3}}\nu \int_{\r}^{\infty}dx\,
\frac{S(x)}{\left(-4  x +  \sinh(4x)\right)^{1/3}}\,,\\
a&=\frac{1}{\cosh(2\rho)}\,.
\label{exactsolution}
\end{aligned}\eeq
where $P_1$ is written in eq.(\ref{lambdaseries}), and $\hat{h}_2=\hat{h}_{KS}$ is the Klebanov-Strassler \cite{KS}
warp factor, as written in eq.(\ref{h2}). The dilaton is a free constant and we take $e^{\Phi}=e^{\Phi(\infty)}=1$. Let us stress that when one takes $S=0$ in eq.\eqref{exactsolution}, 
one obtains precisely the Klebanov-Strassler background.

Serendipitously, 
we write the warp factor as
\beq
\hat{h}= \hat{h}_{KS} + \frac{4}{\epsilon^{8/3}}\nu 
\int_{\r}^{\infty}dx
\frac{S(x) \left(-4  x +  \sinh(4x)\right)^{1/3}}{\left(-4  x +  \sinh(4x)\right)^{2/3}}\,\,.
\label{warpfactorexact}
\eeq
The function
\beq
G(\r)=\int_{\r}^{\infty}dx
\frac{1}{\left(-4  x +  \sinh(4x)\right)^{2/3}}\,,
\label{Green}
\eeq
indicates that
the solution above is just the solution to the Laplace equation
(for the function $\hat{h}$) in the
presence of fluxes $F_5, H_3, F_3$ and D3 sources (see eq.(101) in \cite{KS}).
So, we have the KS geometry being deformed 
by a distribution of source D3-branes
\beq
n_f\sim \nu\,S(\r)\left( \sinh(4\r)-4\r \right)^{1/3}\,,
\label{distributionxxx}
\eeq
that are supersymmetric when placed on the deformed conifold. 
Hence, this exact and analytical 
solution could have been written without going over 
all this effort, just assuming
the  strange distribution of D3 sources in eq.(\ref{distributionxxx}).
On the other hand, the solutions 
with $\l, N_f$ finite --presented in the next section --- are new.
Since the asymptotics of both kinds of solutions are quite similar, we 
can draw generic lessons just by studying 
the exact analytic solutions in eq.(\ref{exactsolution}).
\subsubsection{Generic features of the solutions}
Let us analyze a couple of interesting points.
According to what we wrote in eq.(\ref{distributionxxx}), 
after the rotation, the induced source D3-branes
are distributed in such a way that they pile 
up exponentially towards large values 
of the radial coordinate as $n_f\sim e^{4\r/3}$. This is precisely what 
produces the solutions' departure from the four 
dimensional behavior of the cascade.
This large pile-up of D3-branes dominates the UV dynamics and is equivalent
to the insertion of a dimension-six operator into the Lagrangian, 
as we can see by expanding the warp factor 
$\hat{h}$ in eq.(\ref{exactsolution}) giving as a result the 
one in eq.(\ref{xxyyzz}). We will analyze
later how to get rid of this undesirable behavior.

The second interesting point (as anticipated above)
 is that for a 
suitably chosen distribution $S(\r)$, we can make
the number of sources decrease
towards $\r=0$. In this way, one avoids the presence of the small-$\r$
singularity. 
We can follow the papers \cite{Conde:2011rg}-\cite{Barranco:2011vt} 
to choose an adequate $S(\r)$. 
The backgrounds constructed present no pathology at $\r\sim 0$.
Hence the low-energy strong dynamics of the
field theory I
can be calculated
using the backgrounds above 
and is  fully trustable.
We still need to do something
about the UV behavior, that takes us away from the 
cascade behavior.
Let us make some comments on the field theory dual to our backgrounds
that will illuminate the way to resolving our problem in the large radius region.

\subsection{Field theory comments}

In \cite{Dymarsky:2005xt}
it is explained that 
one may think the moduli space for the quiver 
\beq
SU(N_c+p)\times SU(p)\,,
\eeq
as $(p-l N_c)$ D3-branes free to 
move on the deformed conifold with ($(l+1)N_c$)D5
together with $l N_c$ anti-D5-branes forming a bound state at the threshold.
The ways of distributing the $(p-lN_c)$ D3-branes give
rise to a symmetric product of deformed conifolds as moduli space. 
An important point
made by 
\cite{Dymarsky:2005xt} is that if we start from a quiver 
$SU((k+1)N_c+\tilde{p})\times SU(k N_c +\tilde{p})$ --- in the mesonic branch of 
the
KS field theory ---
and Higgs down the group repeatedly, we will arrive at the quiver
$SU((k+1)N_c)\times SU(kN_c)\times U(1)^{\tilde{p}}$. 
This quiver has the numerology to fit 
the baryonic branch of the KS field theory. 
The  difference is that
the $U(1)_{baryonic}$ is gauged as it mixes with the $U(1)^{\tilde{p}}$.

The authors of \cite{Dymarsky:2005xt} study carefully the 
quantum dynamics of the quiver
in the various ranges of $p$, taking into consideration the presence of
the tree-level and the quantum-induced superpotential.
They show that the 
deformation of the conifold is always proportional to
the scale at which the theory with 
larger gauge group goes strongly coupled.
Let us now apply this information to our case with sources.

Let us consider for the following subsections the case in which the function $S(\r)$ 
vanishes for $\r\leq \r_*$ and stabilizes to $S(\r)\to 1$ for 
large values of the the radial coordinate. Also, let 
us focus our attention on
backgrounds like the one in eq.(\ref{exactsolution}), though the
lessons will be valid also for the semi-analytic solutions 
that we present in Section \ref{semianalyticdetails}.
\subsubsection{ The theory at high energies}
We study the UV of the field-theory dual to the background in
eq.(\ref{exactsolution}).
We interpret this solution
combining the results 
of the papers \cite{Dymarsky:2005xt} and \cite{GMNP}.
We propose that the dual quiver is of the form
\beq
SU(n+ n_f+ N_c)
\times SU(n+ n_f)\,,\;\;\;\; n=k N_c\,.
\label{xaxaxaxa}\eeq
The field theory is in the mesonic branch for $\r>\r_*$, where the
function $S(\r)\sim 1$. 
There are two competing processes, as explained in \cite{GMNP}, 
inspired in early ideas presented  in \cite{hep-th/0101013}.
The usual cascade, represented by the $\hat{h}_{KS}$ 
in the warp factor of eq.(\ref{warpfactorexact})
and a Higgsing process represented 
by the term proportional 
to $\nu$ in eq.(\ref{warpfactorexact}). 
%
In the new radial coordinate $r\sim e^{2\r/3}$,
\beq
\hat{h}|_{\r\to\infty} \sim \frac{\nu r^2+ 3N_c^2 \log r}{r^4}\,,
\label{zaza}
\eeq
we see clearly the superposition of the 
cascade and the Higgsing.
The presence of an exponentially increasing number of
source D3s effectively behaves as the insertion of an 
irrelevant operator of dimension six, deforming the UV dynamics 
out of the near-conformal KS dynamics. 
This indicates the need for a UV completion.
%
\subsubsection{The theory at low energies}
Flowing towards the IR,  
and close to $\r\sim \r_*$ the Higgsing and cascading lower
down the ranks of the groups, so eventually we will 
reach the numerology of the baryonic branch
\beq\bal
& SU(n_f+(k+1)N_c)\times SU(n_f+ kN_c) \to\\
&  SU(n_f+(k+1)N_c -1)\times SU(n_f+ kN_c -1) \times U(1) \to\\
& \to\ldots SU((k+1)N_c)\times SU( kN_c)\times U(1)^{n_f}\,.
\eal\eeq
The main difference with respect to the usual baryonic branch
as discussed in detail in 
\cite{Dymarsky:2005xt} is the fact that the baryon symmetry 
is in this case {\it gauged}. This happens  because 
the baryonic symmetry mixes with a diagonal combination of the $U(1)'s$ 
that appear when Higgsing.
This symmetry being gauged impacts the low-energy 
phenomenology. 
For example, we can check  in the case of the
background in eq.(\ref{exactsolution}), that the 
massless excitation described in
\cite{Gubser:2004qj} is not a solution to the equations of motion.
This implies that the theory is on a  mesonic branch.
 
%
For the rest, with a suitable choice of $S(\r)$
vanishing conveniently fast at $\r=0$, 
the phenomenology of the low-energy field theory is very
similar to that of the Klebanov-Strassler theory, 
as both backgrounds are quite similar. The computation of
various IR quantities will give qualitatively similar results to 
those computed with the backgrounds of \cite{KS} and \cite{BGMPZ}.
There will be numerical differences depending on the profile $S(\r)$.
We will provide examples of this in the following sections.
\subsubsection{How to improve the UV behavior: a phenomenological approach}

\begin{figure}[h]
\begin{center}
\includegraphics[width=0.5\textwidth]{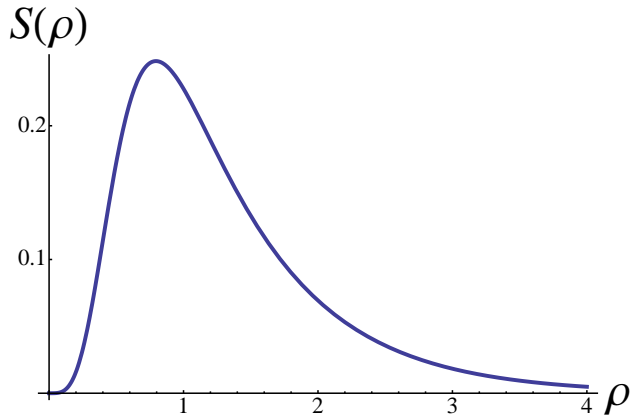}
\caption{Plot of the profile function (\ref{vvvvzzzz}). The function  $S(\rho)$ starts growing for small $\rho$, reaches a maximum, and then decreases exponentially for large values of the holographic coordinate.   }
\label{fig:Srho}
\end{center}
\end{figure}

We would like to stop the growing number of D3 sources
responsible for the pile-up of
Higgsing processes and ultimately for the deviation from the nearly
conformal behavior as 
illustrated in the discussion around eq.(\ref{zaza}).  
A way to do so 
would be to choose a profile for the sources
that somewhat recovers the cascade behavior.
In the case of the background in eq.(\ref{exactsolution}) this is
an easy task because, as we are distributing the D3 sources on the 
deformed conifold, the solution preserves supersymmetry for any $S(\r)$. 
This suggests that we should choose --- at least at a phenomenological level --- a distribution
function behaving as $S(\r)\sim e^{-4\r/3}$ for large values 
of $\r$. A profile that does the job of fixing all the IR and UV problems
is inspired in the work of \cite{Barranco:2011vt} and \cite{Conde:2011rg},
\beq
S(\r)= \tanh(2\r)^4 e^{-4\r/3}\,.
\label{vvvvzzzz}
\eeq
In Figure \ref{fig:Srho} we have plotted the profile function (\ref{vvvvzzzz}). 
By using  this distribution function\footnote{Notice that one is taking $\rho_*=0$ in the example profile \eqref{vvvvzzzz}. It would be trivial to introduce the scale $\rho_*$ by writing something like $S= \Theta(\rho-\rho_*)\tanh(2\r-2\r_*)^4 e^{-4(\r-\r_*)/3}$.}, one can compute the new warp factor using eq.(\ref{warpfactorexact}).
We will then have a smooth IR geometry and a cascade behavior in the far UV.

In principle, one could of course try different profiles that make the distribution of D3-brane
sources either constant or vanishing for large values of $\r$. However, if we want this distribution of sources to have a positive mass density everywhere (and vanishing for $\r\to\infty$), it seems that the only possibility is to have profiles decreasing exactly as $e^{-4\r/3}$. See Appendix \ref{app:EOM} for details.

Another way to motivate why the profile $S(\r)$ should decrease towards the UV\footnote{This argument was suggested
by Anatoly Dymarsky.} is to consider the Klebanov-Strassler \cite{KS} 
geometry with $N_c$ D5-branes 
and add to it $l N_c$ extra D3-branes. The field theory is then in the l-th
mesonic branch \cite{Dymarsky:2005xt} 
of the theory I with groups
$SU((k+l)N_c)\times SU((k+l+1)N_c)$. The relation between the 
deformation parameters $\epsilon_l$ and $\epsilon_0$ of the conifold is
given in \cite{Dymarsky:2005xt}, \cite{Dymarsky:2011pm},
\beq
\epsilon_l= 
\epsilon_0^2\, e^{\frac{2\pi l}{N_c}}
=\epsilon_0^2\, e^{\frac{2\pi l N_c}{N_c^2}}\,.
\eeq
So, we see that 
\beq
N_c^2 \log \epsilon_l \sim l N_c \log\epsilon_0\,.
\eeq
This implies (for similar deformations $\epsilon_l\sim \epsilon_0$), 
that $lN_c\sim N_c^2$. That is, the number of source D3-branes
that we called $n_f$
should be of the same order as the
number of flux D3-branes that we called $n$ in our quiver
of eq.(\ref{xaxaxaxa}). Obviously, 
the situations we discussed above, where the source function $S(\r)\to 1$
at large values of the radial coordinate, generate as explained 
in eq.(\ref{distributionxxx}) an exponentially large number of D3 sources
$n_f$.
Without 
a damping factor like the one proposed in eq.(\ref{vvvvzzzz}) or similar
there is some feature of the field theory that is not 
captured by the solution.
So, we will choose functions describing the profile
of D5 sources (this will directly affect also the profile of D3 sources)
of the form given in eq.(\ref{vvvvzzzz}).
These profiles decrease fast enough in the asymptotic regions, so that
the IR singularity generated by a large density of sources and the UV appearance
of an irrelevant operator are both avoided.

\subsubsection{Further comments on profiles and Goldstones}

To avoid confusion 
with the different uses and meanings of the
function $S(\rho)$,  some words are in order.
The function $S(\rho)$ originated in the supergravity background 
dual to the `field theory II' with D5-brane sources \cite{Conde:2011rg}. 
Its meaning there was that of a profile for 
the D5-brane charge present in the background. 
Coming back to Figure \ref{fig:D5s}, 
we are placing several D5-branes on the geometry, 
each of them reaching a minimal radial distance. 
These minimal distances are distributed on a shell. 
An electrostatic analogue of this situation 
is given  by the electric field created by a 
hollow cavity with a thick charged shell. 
In this case $S(\rho)$ would count the effective radial charge. 
Inside the cavity $S(\r)=0$ as there is no electric field. 
As we cross the shell, $S(\rho)$ increases, 
and stabilizes to the total charge away from the shell. 
The electric field outside the shell,
will be that of a point charge. Had the shell
null thickness, the electric field would display a jump. 
This is what the $S(\rho)$ in Figure \ref{fig:D5s} represents. 
Away from the shell, $S(\rho\to\infty)\sim 1$
in coincidence with the $S(\rho)$ generated by `massless' sources.

This interpretation of $S(\rho)$ as a profile for the D5-brane charge 
is however lost after the rotation. 
We gave an interpretation of $S(\rho)$ for the 
exact solution of eq.\eqref{exactsolution}. As explained
after eq.\eqref{Green}, in this new solution $S(\rho)$ 
accounts for how the 
D3-branes are distributed in the geometry. 
So, for this exact-analytic solution, $S(\rho)$ represents a distribution 
of D3-branes. More precisely, 
the distribution is given by $S(\rho)\left(\sinh(4\r)-4\r\right)^{1/3}$.

A curious fact is that the 
`physically sensible' $S(\r)$ for the unrotated background 
(which asymptotes as $S\sim 1$), is pathological for the 
rotated configuration as the number of D3-branes grows 
exponentially with $\rho$, inducing an irrelevant deformation. 
Conversely, 
a physically sensible $S(\rho)$ in the 
rotated background (vanishing at infinity as $e^{-4/3\rho}$) 
does not have --- at present --- 
a microscopic  
interpretation in terms of the usual 
flavor D5-branes in the unrotated solution. 
A clarification of this point remains for future work.

%
%
{\it About Goldstone bosons:}
Let us briefly discuss the situation with Goldstone bosons. 
Detailed calculations needed
to sort out the issue in full are left for  
future work\footnote{Many thanks to Anatoly Dymarsky
for crucial comments reflected in these following paragraphs.}.

First, let us state the question.
Consider a background like the one in eq.(\ref{exactsolution}).
We can choose any $S(\r)$ with a similar behavior 
to that in eq.(\ref{vvvvzzzz}), without affecting the argument.
In practical terms, we are making a transition from a 
far UV with similar dynamics to
KS and the baryonic branch, to an intermediate region, with 
the numerology of the mesonic branch, to a deep IR which 
has the gauge-group ranks of the baryonic branch of \cite{KS}. 
If one then calculates the solution for the 
massless mode in~\cite{Gubser:2004qj}, 
the proposed ansatz of~\cite{Gubser:2004qj} 
does not verify the linearized equations of motion. 
Indeed, despite the fact that $F_3$ and $B_2$  
in eq.(\ref{exactsolution}) are the same as in the KS background, 
the five-form flux $F_5$ is modified by the presence 
of sources (as seen from the fact that $\hat h \neq \hat h_{KS}$). 
Then, in particular, the equation of motion 
for $H_3$ would be different from the standard KS one, 
including a term involving $dF_5$ which is non-zero because of the sources.

In line with this, as we emphasized above, transitions
between mesonic and baryonic branches should 
produce gauged baryonic symmetries \cite{Dymarsky:2005xt}, and as such
the Goldstone of the baryonic $U(1)$ should 
not be part of the physical spectrum
(it should be Higgsed away into the resulting massive gauge boson).

Nevertheless, one may speculate that a modification
of the ansatz in \cite{Gubser:2004qj} may reveal the existence
of a massless mode.
If we were to find a 
Goldstone boson on the gravity side, 
how could we understand it in terms of the field-theory dual?
In the combination of Higgsing and cascading,
there is a complicated pattern of global and gauge symmetries, 
which are non-trivially mixed with one another, and which undergo partial
symmetry breaking. Vacuum-alignment arguments suggest that,
if possible (i.e.~in the absence of non-trivial obstructions
within the Lie groups and 
cosets describing the symmetry of the system), the vacuum of the theory 
will align itself with the directions 
in the internal space of the global symmetry. 
In doing so, the vacuum will
prefer to spontaneously break part of the symmetry which is {\it not} gauged.
As we wrote, a detailed calculation would be 
needed in order to establish whether or not there is 
a normalizable massless mode  in the spectrum,
and we leave this for a future study.

We move now to the study of the
semi-analytic solutions described in the beginning of Section \ref{twokinds}.
\section{Details on the new solutions}\label{semianalyticdetails}
In this section we  give details for the semi-analytic solutions. We will 
concentrate
on the configuration 
after the rotation. We consider the case in which the parameters $N_f,\l$
are free. While the solutions could be written in terms of an infinite series
of integrals (that cannot be performed exactly), we find it more useful  
to
give series expansions close to $\r=0$ and $\r\to\infty$
and a numerical interpolation between both regimes. 
After  solving the master equation (\ref{master-eq}), we will
use  the expressions in eqs.(\ref{h-k-P-Q})-(\ref{bderho})
to quote the asymptotic expansions for the functions appearing in
the background.
As anticipated in 
eq.(\ref{vvvvzzzz})
 we will consider (this is just a possible example) 
the function $S(\r)= \tanh^4(2\r)e^{-4\r/3}$.

We should clearly state that
we are not obtaining the profile above in a  rigorous fashion.
Since the distribution of D3-branes on the deformed conifold can be chosen
(they  are SUSY), 
the solutions with $\l\to\infty, N_f\to 0$ of eq.(\ref{exactsolution}) are
correct. The ones we present in this section, lack 
a rigorous
proof of existence (we are choosing a profile of D5-branes that we did 
not prove to be derived from a kappa-symmetric embedding). The 
healthy behavior
of the backgrounds we found (together with the positivity
of $T_{00}$ for the sources) suggests that the brane embedding can 
be rigorously derived.
Let us start by studying the large-$\r$ expansion of the solutions
to the master equation (\ref{master-eq}) given the source term in 
eq.(\ref{vvvvzzzz}).

\subsection{Large-$\r$ expansions}

First of all we define
\beq
S_\infty=\int_{0}^{\infty}d\rho\,\tanh^2(2\r) S(\rho)\,,
\eeq
so that we can write for the asymptotic behavior of the function $Q(\r)$:
\beq
Q=N_c(2\r-1) - 
N_f S_\infty +\frac{3}{4}N_f e^{-4\r/3} +
(4N_c \r -2 N_f S_\infty)e^{-4\r}+\co(e^{-16\r/3})\,.
\eeq
The function $P(\r)$ is given at large $\r$ by
\beq\bal
P(\r)&= e^{4\r/3}\Bigg( c_+   +\frac{N_c^2}{c_+} e^{-8\r/3}\left(4
\r^2 -4\r(1+\frac{N_f}{N_c} S_\infty)+
\frac{13}{4}+ \frac{3 c_+ N_f}{4N_c^2} + 2 \frac{N_f}{N_c} S_\infty + \frac{N_f^2}{N_c^2} S_\infty^2
\right)+\\
&\quad + \frac{e^{-4\r}}{3c_+}\left( 2 N_c N_f \r^2 -2\r
(4c_+^2 - N_c N_f + N_f^2 S_\infty) \right) + \co(e^{-16\r/3}\r^2) \Bigg)\,.
\label{Puvxx}
\eal\eeq
Using eqs.(\ref{h-k-P-Q})-(\ref{bderho}), we get 
for the background functions
\begin{align}
\hat{h}=& 3\frac{e^{-8\r/3}}{8c_+^2}
\left(N_c^2(8\r-1)+2 c_+ N_f - 4 N_c N_f S_\infty     \right)
- \frac{N_f}{2c_+^2}e^{-4\r}\big( 2N_c(\r-1)- N_f S_\infty \big)+\nonumber\\ 
&+ \co(e^{-14\r/3})\,,\nonumber\\
e^{2h}=& \frac{c_+}{4}e^{4\r/3} +
\frac{N_c(2\r-1)- N_fS_\infty}{4} 
+\frac{e^{-4\r/3}}{16c_+}\Big(
16 N_c^2 \r^2 -16 N_c(N_c+N_fS_\infty)\,\r\, +\nonumber\\
&+ 13 N_c^2 + 6 c_+ N_f + 8 N_c N_f S_\infty + 4 N_f^2 S_\infty^2  \Big)
+ \co(e^{-8\r/3})\,,\nonumber\\
\frac{e^{2g}}{4}=& \frac{c_+}{4}e^{4\r/3} 
-\frac{N_c(2\r-1)- N_fS_\infty}{4}+\frac{e^{-4\r/3}}{16c_+}
\Big( 16 N_c^2 \r^2 -
16 N_c(N_c+N_fS_\infty)\,\r \,+\nonumber\\   
&+ 13 N_c^2 + 8 N_c N_f S_\infty + 
4 N_f^2 S_\infty^2  \Big)+ \co(e^{-8\r/3}\r^2)\,,\nonumber\\
\frac{e^{2k}}{4}=& \frac{c_+}{6}e^{4\r/3}-\frac{e^{-4\r/3}}{24c_+}
(4 N_c \r - 5 N_c -2 N_f S_\infty)^2+\nonumber\\
&+e^{-8\r/3}\frac{(4c_+^2+N_fS_\infty)(8\r-3)-N_cN_f(8\r^2+2\r-3)}{36c_+}+ \co(e^{-4\r})\,,\nonumber\\
a=& 2 e^{-2\r}+2e^{-10\r/3}\frac{N_c(2\r-1)- N_f S_\infty}{c_+}+\nonumber\\
&+e^{-14\r/3}\frac{4N_c^2(2\r-1)^2-8N_cN_fS_\infty(2\r-1)+4N_f^2S_\infty^2+3c_+N_f}{2c_+^2}+\cO(e^{-6\r})\,,\nonumber\\
N_c b=& e^{-2\r}(4N_c \r -2 N_f S_\infty) 
+\frac{N_f}{2}e^{-10\r/3}+e^{-14\r/3}\left(4N_c \r -2 N_f S_\infty\right)+ \co(e^{-6\r})\,,\nonumber\\
e^{4\Phi}=&e^{4\Phi(\infty)} \bigg(
1 - \frac{3e^{-8\r/3}}{4c_+^2}\left[N_c^2(8\r-1)+2 c_+ N_f - 
4 N_c N_f S_\infty     \right]+ \nonumber\\
&+\, \frac{N_f}{c_+^2}e^{-4\r}\big( 2N_c(\r-1)- N_f S_\infty    \big) \bigg)
+ \co(e^{-14\r/3})\,,
\label{semianalyticUV}
\end{align}
where $e^{4\Phi(\infty)}=\frac{3e^{4\Phi_0}}{8c_+^3}$. Notice that one effect of the sources --- with this given profile $S(\r)$ ---
is to add to the D3-brane charge in the warp factor.
The sources do not interfere with the cascade, that is 
now driven by the bulk D3-branes only,
associated with the number $n$ in our quiver.
Let us discus the IR expansion next.

\subsection{Small $\r$ expansions}

\begin{figure}[h]
\begin{center}
		\includegraphics[width=\textwidth]{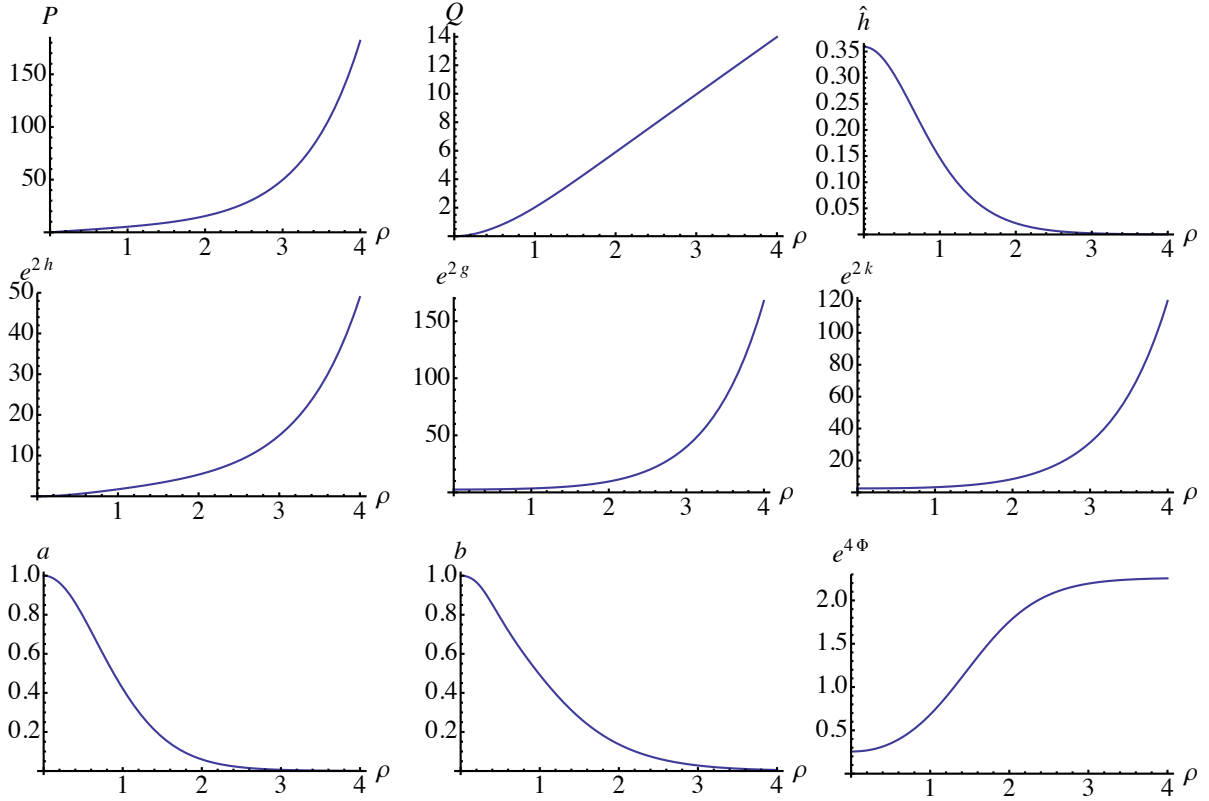}
\end{center}
\caption{For the case $N_c = N_f =2$ and $h_1 = 5$, we show 
the functions defining
the semi-analytic rotated background. The choice of $S(\r)$ is the
one specified in the text (eq. (\ref{vvvvzzzz})).}
\label{fig:tutti}
\end{figure}

With the same $S(\r)$ as  in (\ref{vvvvzzzz}), we can see that close to $\r=0$, 
the effect of the sources vanish as
\beq
S(\r)= 16 \r^4 -\frac{64}{3}\r^5 -\frac{640}{9}\r^6+\cO(\r^7)\,.
\eeq
It is clear that the deep IR asymptotics will suffer 
little modifications from that of
the case with $N_f=S(\r)=0$.
In order to keep track of the IR modifications, we just need to look at the
terms with a factor $N_f$ and compare the expansions with those
in eq.(\ref{P-IR.A}) and eqs.(\ref{SUSYIREXPapen}).
The functions $Q(\r), P(\r)$ read
\bea
& & Q(\r)= \frac{4N_c}{3}\r^2 -\frac{16N_c}{45}\r^4 +(\frac{128N_c}{945} 
-\frac{32N_f}{7})\r^6+\cO(\r^8)\,,\\
& & P(\r)= h_1\r +4\frac{h_1^2 - 4N_c^2}{15h_1}\r^3 +\frac{16}{1575 h_1^3}
(3h_1^4 -4 h_1^2 N_c^2 - 32 N_c^4 - 450h_1^3 N_f )\r^5+\cO(\r^7)\,,\nonumber
\eea
while the functions in the background have expressions given by\footnote{
Notice that for the function $a(\r)$ above, the first correction of the sources 
appears at order $\r^6$.
}
\begin{align}
\hat{h}&=1-\frac{e^{2\Phi(0)-2\Phi(\infty)}}{2\sqrt{2}} -\frac{8\sqrt{2}N_c^2}{9h_1^2}e^{2\Phi(0)-2\Phi(\infty)} \r^2-\NO\\
&\quad-\frac{2\sqrt{2}}{135}e^{2\Phi(0)-2\Phi(\infty)}\left(-40\frac{N_c^2}{h_1^2}+224\frac{N_c^4}{h_1^4}+135\frac{N_f}{h_1}\right) \r^4+\cO(\r^5)\,,\NO\\
e^{2h}&=\frac{h_1 \rho ^2}{2}+\frac{4}{45} \left(-6 h_1+
15 N_c-\frac{16 N_c^2}{h_1}\right) \rho ^4+\NO\\
&\quad+\frac{16}{4725h_1^3}\!\left(267 h_1^4 +1814 h_1^2 N_c^2 -
700 h_1 N_c^3 -328 N_c^4 -15h_1^3(77 N_c + 45 N_f)\right)\r^6 +
\co(\r^7)\,,\NO\\
\frac{e^{2g}}{4}&=\frac{h_1}{8}+
\frac{1}{15} \left(3 h_1-5 N_c-\frac{2 N_c^2}{h_1}\right) \rho ^2+\NO\\
&\quad+\,\frac{2 \left(3 h_1^4+10 h_1^3 (7N_c-45N_f)-144 
h_1^2 N_c^2-32 N_c^4\right) \rho ^4}{1575 h_1^3} +\co(\r^5)\,,\NO\\
\frac{e^{2k}}{4}&=\frac{h_1}{8}
+\frac{\left(h_1^2-4 N_c^2\right) \rho ^2}{10 h_1}
+\frac{\left(6 h_1^4-8 h_1^2 N_c^2-64 N_c^4 - 270h_1^3 N_f\right) \rho ^4}{315 
h_1^3}
+\co(\r^5)\,,\NO\\
e^{4\Phi}&=e^{4\Phi(0)}\left(1+\frac{64  N_c^2 \rho ^2}{9 h_1^2}
+\frac{16( -120 N_c^2h_1^2 +992N_c^4 +405 h_1^3 N_f ) \rho ^4}{405 h_1^4}+
\co(\r^5)\right)\,,\NO \\
a &= 1+\left(-2+\frac{8 N_c}{3 h_1}\right) \rho ^2
+\frac{2 \left(75 h_1^3-232 h_1^2 N_c+160 h_1 N_c^2+64 N_c^3\right) 
\rho ^4}{45 h_1^3}
+\co(\r^6)\,,\NO\\
b &=  1-\frac{2}{3}\r^2 
+(\frac{14}{45} -\frac{8N_f}{N_c})\r^4  +\co(\r^5)\,,
\label{SUSYIREXP}
\end{align}
where $e^{4\Phi(0)}=\frac{8e^{4\Phi_0}}{h_1^3}$. Using these expressions, one can find a smooth numerical interpolation between 
the two asymptotics and  compute observables with them.
In  Figure \ref{fig:tutti}, we plot the functions $P(\r)$ and $Q(\rho)$ 
for $S(\r)=\tanh^4(2\r)e^{-4\r/3}$. Using $P(\r), Q(\r)$ and $S(\r)$
all other functions in the background can be obtained.
Studying these functions, we do not observe 
any pathology or non-physical behavior. Hence, we consider 
the choice of $S(\r)$
in this section to be  good.

\section{Further analysis of field-theory quantities}\label{QFTQD}

In this section, we continue the study of the background found 
in Section \ref{semianalyticdetails}. 
We compute some quantities in the proposed dual field theory.
The purpose of these calculations is to investigate if the profile chosen
for $S(\r)$ produces any undesirable feature that might
suggest non-physicality
of our arbitrary choice. We start with the central charge in the dual QFT.
\subsection{Central Charge}
\begin{figure}[h]
\begin{center}
		\includegraphics[width=0.5\textwidth]{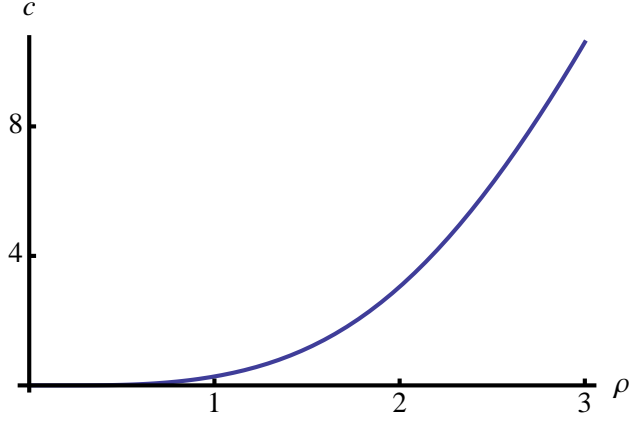}
\end{center}
\caption{Plots of the function $c(\r)$ for $N_c = N_f =2$ and $h_1 = 5$. 
We considered the backgrounds
of Section \ref{semianalyticdetails}.}
\label{fig:RotatedCentralCharge}
\end{figure}

We calculate the central charge $c$, following the prescription developed in
\cite{CentralCharge}, \cite{Klebanov:2007ws}.
Start with a metric (in string frame) of the form
\beq
ds_{st}^2= \a\, dx_{1,d}^2 +\a \beta\, d\r^2 + g_{ij}\,dy^i dy^j\,\,,
\eeq
and following \cite{Klebanov:2007ws}
we define $V_{int}$ as the volume of the internal space, and $H$ as
\beq
H= e^{-4\Phi}V_{int}^2\, \a^d\, .
\eeq
We have that the central charge (for $d=3$) is
\beq
c=\frac{\beta^{3/2}H^{7/2}}{(H')^3}\,.
\eeq
In particular, for a background like in eq.(\ref{configurationfinal}),
we have
\beq
\bal
\a&=e^{\Phi}\hat{h}^{-1/2}\,, \;\;\; 
\beta=e^{2k}\hat{h}\,,\;\;\; V_{int}^2= e^{4h+4g+5\Phi+2k}\hat{h}^{5/2}\,,\;\;\;
H=e^{4\Phi+4g+4h+2k} \hat{h}\,, \\
c&= \frac{\hat{h}^2e^{2\Phi+2h+2g+4k}}{8(\partial_\r \log[\sqrt{\hat{h}}
e^{2\Phi+2h+2g+k}])^3}\,.
\label{centralafter}
\eal
\eeq
We consider the semi-analytic backgrounds of Section \ref{semianalyticdetails} 
and plot the central charge of eq.(\ref{centralafter}) 
in Figure \ref{fig:RotatedCentralCharge}. We observe 
a monotonic behavior. This quantity does not show 
any sign of unphysical behavior for the chosen profile\footnote{One could have arbitrarily chosen a different profile, for example
$S\sim \tanh^4(2\r)e^{-(\r-\r_0)^2}$. In this case we would 
have observed a non-monotonic $c$-function.}.

For backgrounds like those in eq.(\ref{f3old}) ---
before the rotation --- we should 
take $\k=0$ and $\hat{h}=1$ in the formulas above.

\subsection{Quantities defining the gauge groups}
In specifying our quiver, for example around eq.(\ref{xaxaxaxa}), 
we have introduced various quantities: $n$, that is the number 
of bulk (flux) D3-branes; $n_f$, that is the number of source D3-branes;
$N_c$ and $N_f$, the numbers of bulk and source D5-branes.
Here, we give a geometric definition for these quantities, forgetting about numerical factors.
\beq
	\begin{aligned}
		n + n_f &= \int_{\Sigma_5} F_5 = \int_{\Sigma_5} B_2 \wedge F_3\,\,, \\
		n &= \left(\int_{\Sigma_5} F_5 \right)\! \bigg |_{N_f = 0}\,,\;\;\;\;\qquad
n_f = \int_{\Sigma_5} F_5 - \left(\int_{\Sigma_5} F_5 
\right)\! \bigg |_{N_f = 0}\,\,,
	\end{aligned}
\eeq
where $\Sigma_5$ is the five-dimensional manifold of coordinates $(\theta, \varphi, \tilde\theta, \tilde\varphi, \psi)$.
Using the expressions in eq.(\ref{configurationfinal}), 
we can calculate $F_5\big|_{\Sigma_5}$ to be
\beq
	\begin{aligned}
F_5\big |_{\Sigma_5}=	B_2 \wedge F_3 = - \frac{\kappa}{8} e^{2\Phi} \Bigg[ &\frac{\cosh(2\r) -a}{\sinh(2\r)} \left( -2N_c e^{2h} + \frac{N_c}{2} e^{2g} (a^2 -2 a b +1) - \frac{N_f}{2} e^{2g} S\right)- \\
	&- \frac{4 N_c}{\sinh(2\r)} e^{2h} (a-b) \Bigg] \sin \theta \sin \tilde \theta d\theta \wedge d\varphi \wedge d\tilde \theta \wedge d\tilde \varphi \wedge d\psi\,,
	\end{aligned}
\eeq
that translates into
\beq
	\begin{aligned}
		n + n_f &= 8 \pi^3 \kappa\, e^{2\Phi} \bigg[ \frac{\cosh(2\r) -a}{\sinh(2\r)} \left( 2N_c e^{2h} - \frac{N_c}{2} e^{2g} (a^2 -2 a b +1) + \frac{N_f}{2} e^{2g} S\right)+ \\
		&\qquad\qquad\quad+ \frac{4 N_c}{\sinh(2\r)} e^{2h} (a-b) \bigg]\,,\\
		n &= (n+n_f)|_{N_f =0}\,, \;\;\;\qquad
n_f = (n+n_f) - (n+n_f)|_{N_f =0}\,.
	\end{aligned}
\eeq
While it is good to have the exact expressions above, for many purposes
the large and small $\r$ expansions are illuminating.
For low energies in the dual QFT, we have
\beq
	\bal
		n &\propto N_c^2\r^3 + \left(\frac{32N_c^4}{9h_1^2}-\frac{4N_c^2}{5}\right) \r^5 +\cO \left(\r^7\right)\,\,, \\
		n_f &\propto N_f h_1\left[\r^5 -\frac{4}{3} \r^6 + \cO\left( \r^7 \right)\right],
	\eal
\eeq
which makes it clear that deep in the IR the Higgssing (represented by $n_f$)
has stopped. The geometry is very similar to that of
\cite{BGMPZ} and so is the dual QFT. To describe the regime of high energies 
in the field theory, we change to a radial coordinate $r=e^{2\r/3}$ as defined around
eq.(\ref{zaza}). Then,  for large $r$ one has
\beq
	\bal
		n &\propto N_c^2 \frac{3 \log r -1}{3} - N_c^4 \frac{36 \log^2 r -15 \log r +1}{8 c_+^2 r^4} + \cO \left(r^{-6}\right) \,\,,\\
		n_f &\propto N_f \left[\frac{c_+ -2N_c S_\infty}{6} - \frac{6N_c \log r-5N_c -2N_f S_\infty}{12r^2} + \cO \left(r^{-4}\right)\right].
	\eal
\eeq
We see that far in the UV the cascade is at work, with a 
number of `bulk' D3-branes $n=k N_c$ --- as usual $k\sim N_c \log(r)$.
The Higgsing has stopped and the source D3-branes $n_f$ just accounts
for a constant charge. This situation is also reflected in 
the warp factor $\hat{h}$, see the discussion around 
eq.(\ref{semianalyticUV}).

\section{Towards applications.}\label{SECTIONAPP}

This section is devoted to provide  the reader with a reason,
based on field-theory arguments, why this type of solutions
is interesting and might be relevant  for phenomenology and model-building 
in high-energy Physics.
In terms of the field theory dual, 
the backgrounds constructed and discussed in this paper have a set of very 
peculiar features,
which can be summarized as follows.

First of all, notice that (besides the confinement scale related to the 
end of space in the geometry)
  there are three dynamical scales in the theory. 
One is the scale $\bar{\r}$ controlled by the parameter
$c_+$ in eq.(\ref{UV-II-N}), i.e.~by the insertion of a 
dimension-eight operator
in the single-site theory
(which is also related to a dimension-two VEV, as explained 
in~\cite{Elander:2011mh} and references therein).
In the absence of sources, below such scale the theory is best described as 
a generalization
of the one-site field theory dual to~\cite{MN}, while above this scale the 
rotation procedure
yields a background that is dual to a two-site quiver realizing the 
(baryonic branch of the) 
Klebanov-Strassler duality cascade.
There are then two other scales $\rho_*$ and $\rho_S$, $\rho_*<\rho_S$, such that the function $S$ 
has support in the range
between them.
There is no obvious relation between $\bar{\r}$ and these other two scales,
but for reasons of simplicity we will assume in this discussion that 
$\bar{\r}\simeq \r_S$.
Different cases can be discussed along the same lines, in a 
case-by-case 
way that 
does not add to the main physical points we want to make.

We will focus only on the rotated solutions.
\begin{itemize}
\item In the far UV, for $\r>\r_S\sim \bar{\r}$,
 the theory resembles the (baryonic branch of) the KS cascade:
the theory is flowing close to a line of fixed points, 
each of which is a ${\cal N} =1$ Klebanov -Witten fixed
point,  a two-site quiver with gauge group $SU(n)\times SU(n)$, and with $n$ 
increasing towards the UV
(i.e.~with $\r$). The flow never approaches an actual fixed point at finite 
(and large) $n$ 
because of a small imbalance between the ranks of the two gauge groups, but 
rather the flow goes up a cascade of Seiberg dualities \cite{Seiberg:1994pq}
which continues 
indefinitely towards $\r\rightarrow +\infty$. See \cite{Strassler:2005qs}.

\item There is an intermediate range $\r_{\ast}<\r<\r_{S}\sim\bar{\r}$ over 
which 
the function $S$ is non-trivial. At the scale $\bar{\r}$ 
and in the absence of sources, the duality cascade would stop, 
due to the Higgsing induced by the dimension-two condensate appearing and 
precipitating the theory towards the last stages of the duality cascade itself~\cite{GMNP,Elander:2011mh}.
On the other hand, because in this range the function $S$ is non-trivial,
 another cascade starts, which has a completely different interpretation:
 it is  a cascade of Higgsings of the gauge theory. 
Source D3s are crossed when flowing to the IR, hence realizing
the ideas proposed in
\cite{hep-th/0101013}.

\item Below the  scale controlled by the value of $\r_{\ast}$, the Higgsing cascade
stops,
and with it most of the dynamical features related to  $N_f$ (up to subtleties
which have been discussed earlier), because $S$ 
vanishes.
The theory now looks like a generalization of the single-quiver theory 
in~\cite{MN},
with the same type of dynamics.

\item At very low scales (near the end of space) of the geometry,
the theory shows the appearance of a non-trivial gaugino condensate, and 
confines in the usual sense of producing an area law for the Wilson loop.
\end{itemize}
Notice that one might as well take $\bar{\r}$ to be very small, near the end of 
space:
in this case the duality cascade would continue all the way to very small 
energies,
in particular extending over the second and third of the four ranges described 
above.
In this case, there would be a regime in which both Higgsing cascade and duality 
cascade coexist.

The physics taking place in the field theory at the second of the four stages
(for  $\r_{\ast}<\r<\r_S\sim\bar{\r}$)
is very peculiar, and even more peculiar and new is the fact that such a behavior 
appears only over a finite range of energies.
The question is: what kind of theoretical
models  exhibit features that are qualitatively similar to the one discussed 
here, 
and what kind of possible physical systems can they describe?
In other words: are there conceivable applications of such results?

Interestingly, there is a positive answer to this question, which relates
to a long-standing, very difficult and open phenomenological  problem.
A scenario that has some common elements with our present one is that of
Extended Technicolor Models (ETC) with Tumbling dynamics~\cite{ETC}.
This is a very plausible dynamical explanation for 
the origin of flavor physics (SM-fermion masses, mixing angles,
CP violation, FCNC interactions, \dots), in the context of strongly coupled 
extensions of the Standard Model~\cite{TC,WTC} (see also~\cite{reviews} for a sample of 
reviews on the subject).
We will  provide here a simple summary of what this means, 
besides referring the reader to the literature on the subject.
We also want to clarify what are the actual similarities
and the substantial differences, and hence the possible
intrinsic limitations into attempting to use the approach of this paper in order 
to study tumbling ETC.

 ETC addresses the following, well-known,  fundamental problem, arising in the 
context of 
 (strongly coupled) Dynamical ElectroWeak-Symmetry Breaking (DEWSB). DEWSB
or technicolor (TC) proposes to replace the Higgs sector of the Standard Model 
with 
a new gauge theory with group $G_{TC}$ and new (techni-fermion) fields 
transforming non-trivially
under the action of both $G_{TC}$ and the Standard Model gauge group.
The strong dynamics associated with $G_{TC}$ yields the formation (at the 
confinement scale $\Lambda_{TC}$) 
of non-trivial condensates made of techni-fermions, which results in the 
spontaneous breaking of the 
SM gauge group at the ElectroWeak scale. The SM gauge interactions themselves 
communicate
the breaking to the $W$ and $Z$ gauge bosons, which become massive.
However, in the process of removing the Higgs field, one also
 loses the Yukawa couplings,
and hence one needs a mechanism that couples the SM fermions to the techni-quarks,
in order for the quarks and leptons to acquire a mass below the ElectroWeak 
scale.

\begin{figure}[h]
\begin{center}
\begin{picture}(300,200)
\put(10,10){\includegraphics[height=7cm]{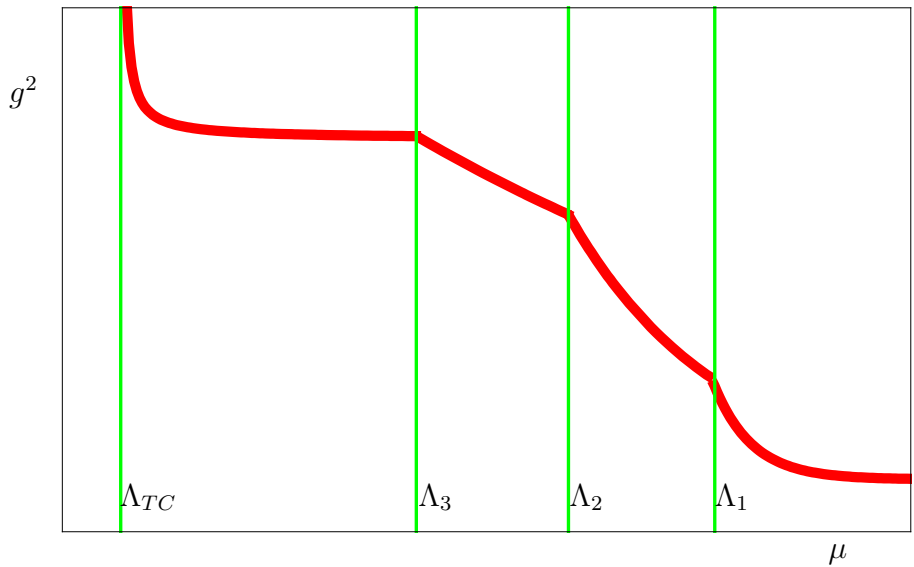}}
\put(-10,173){${g^2}$}
\put(257,20){$\Lambda_1$}
\put(202,20){$\Lambda_2$}
\put(145,20){$\Lambda_3$}
\put(32,20){$\Lambda_{TC}$}
\put(300,0){$\mu$}
\end{picture} 
\caption{Cartoon depiction of the running of the gauge coupling $g^2$ in an example of
multi-scale (tumbling) ETC model as a function of the renormalization scale $\mu$.
The sequential breaking at scales $\Lambda_1>\Lambda_2>\Lambda_3>\Lambda_{TC}$
means that the gauge group and field content changes at each breaking scale, and so does the dynamics.
A model that is believed to exhibit such behavior is described in 
detail in~\cite{APS}.}
\label{Fig:ETC}
\end{center}
\end{figure}

ETC provides such a dynamical mechanism.
The generic ETC model works in the following way (an example of
such a construction is discussed in detail
in~\cite{APS}).
Start from some gauge theory with gauge group $G_{ETC}\times G_{SM}$ (where 
$G_{SM}=SU(3)_c\times SU(2)_L\times U(1)_Y$
is the familiar SM gauge group),
and  a given fermionic-matter content.
Assume that the dynamics of $G_{ETC}$ is such that the theory is asymptotically 
free,
but undergoes a sequence of breaking stages
\beqs
\nonumber  
G_{ETC}\rightarrow G_1\rightarrow G_2
 \rightarrow G_{TC}\,,
 \eeqs 
 at scales $\Lambda_1\gg\Lambda_2\gg\Lambda_3$, respectively.
 At scales below $\Lambda_3$ the resulting effective field theory consists 
of the following. 
 \begin{itemize}
 \item A gauge theory with gauge group $G_{TC}\times G_{SM}$.
 \item Two kinds of massless fermions: techni-quarks ${\psi}_{TC}$ transforming 
non-trivially 
 under $G_{TC}\times G_{SM}$,
 and singlets of 
$G_{TC}$ 
  that we denote by ${\psi}_{SM}$,  
that transform non-trivially only under 
$G_{SM}$. 
 The latter are identified with the quarks and leptons of the Standard Model.
 \item Higher-order operators originating from integrating out the 
heavy gauge bosons 
 of the coset $G_{ETC}/G_{TC}$. In particular, some of these are four-fermion 
operators
 coupling two SM fermions with two TC fermions,
 with the generic form 
 \beqs
 \frac{1}{\Lambda_i^2}\bar{\psi}_{TC}{\psi}_{TC}\bar{\psi}_{SM}{\psi}_{SM}\,.
 \eeqs
 \end{itemize}
 We present in Figure \ref{Fig:ETC} a cartoon of the dynamics (in terms of the effective gauge coupling)
 of such a scenario.
 
 Now, the resulting gauge theory is the TC theory. 
 Ultimately, $G_{TC}$ will confine, and produce condensates that break $G_{SM}$.
 However, the presence of the four-fermion interactions
 means that after ElectroWeak-Symmetry Breaking the quarks and leptons will 
become massive.
 Effectively, these four-fermion operators play the same role as the Yukawa 
couplings
 in the Standard Model. Notice that, because they originate at different scales,
 there will be in general three families of SM fermions, and 
 some of these operators will be suppressed as $1/\Lambda_1^2$, others as 
$1/\Lambda_3^2$,
 and so on.
 In general there will be a very complicated, hierarchical structure in the 
four-fermion
 couplings, which 
 will translate after dimensional transmutation into the  hierarchical 
 structure of the masses of the SM fermions, and the mixing angles in the CKM 
mixing matrix.

 The presence of a long (at least three-stage) sequence of breaking,
 and of hierarchies in the dynamical scales is absolutely necessary on 
phenomenological grounds,
 because it offers the only plausible and self-contained
  dynamical explanation for the pattern of 
 phenomenological masses and mixing angles experimentally measured.
 On the other hand, it makes it extremely difficult to study, because of the 
strongly coupled nature of the 
 phenomena taking place in these field theories.
Many open questions about such dynamics, and its low-energy implications, 
require some new tools for a quantitative (and often even qualitative)
 analysis to be performed.
The gravity duals we are finding resemble this scenario: a sequence of Higgsing 
stages
taking place at several scales, over a finite energy range.
It is hence conceivable that some of the long-standing open problems might be 
addressed in this context.
For example, one would also like to understand what kind of low-energy 
spectrum one may observe, with particular reference to the presence of possible 
pseudo-Goldstone bosons associated with the
breaking of accidental global symmetries.

There is another long-standing problem (see for instance~\cite{AW} and references therein). 
Besides the gauge symmetries, all of the
TC and ETC models also possess very large (approximate) global symmetries,
which are spontaneously broken by the many condensates
that form. This might yield the presence of such pseudo-Goldstones,
with masses that might put them well below the current exclusion region, and 
hence
render the models phenomenologically not viable. Computing masses and couplings of such particles requires dedicated strong-coupling
calculations, and new tools are needed to perform them.

We can hence conclude that this is a first concrete example of a model which 
shares some
of the fundamental features of {\it tumbling}, and that these types of models 
might be used to characterize 
in field-theory terms what are the features associated with the tumbling itself.
It must also be stressed that the very nature of these models is such that a 
direct comparison to 
the real world must be done with caution:  effectively we have an
infinite, continuum number of Higgsing stages, rather than a few (three) distinct 
and hierarchical breaking stages.
In other words, comparing to Figure \ref{Fig:ETC},
the Higgsing cascade differs from the tumbling because between the 
scales $\Lambda_1$ and $\Lambda_3$ (which can be associated with $\r_S$ and $\r_{\ast}$, respectively),
one has a continuum of breaking scales, rather than just a few steps.
 It hence remains to be understood what type of 
operators will replace the four-fermion operators 
of low-energy TC, and what type of phenomenological implications they have.
This is somewhat analogous to the fact that the gravity dual of the duality cascade 
actually consists of an
infinite continuum of Seiberg dualities, rather than a small number of such 
stages.

\section{Conclusions}\label{sectionconclusions}
Let us conclude this paper by recalling the main results of this work and
proposing some  topics for future investigation.
The reader who wants to have a simple summary 
of the results and main original features discussed in the paper
 should complement this final section with the brief summary at the beginning of 
Section \ref{SECTIONAPP}.

The basic idea of this paper is to construct new classes of 
gravity backgrounds that we interpret in terms of the very non-trivial  RG flow of
a dual field theory with ${\cal N}=1$ supersymmetry,
which exhibits a number of interesting properties.

We  constructed these new solutions, which generalize the KS~\cite{KS}
and baryonic branch~\cite{BGMPZ} solutions, by the addition of sources 
into the type IIB configuration. 
We proposed suitable profiles for the sources that
avoid any small radius singularity and also provide the background with a
`logarithmically' AdS space in the large-radius regime. In this fashion, 
the backgrounds are trustable
and the dual field theory associated is well defined all along the flow.
It is therefore sensible to think that these backgrounds are characterizing a 
well-defined field theory, and hence we suggested an interpretation 
for the microscopic properties of such a field theory.
We studied different aspects 
of the four-dimensional dual quantum field theory that point 
to the consistency of the interpretation given.

A comment on the string side of the construction must be made.
While the solutions with only D3 sources on the deformed conifold are 
fully rigorous, our profiles for D5 and D3 sources have not been obtained
from a microscopic kappa-symmetric solution 
to the brane equations of motion.
In spite of this, various tests suggest that the backgrounds 
presented here are physically acceptable, both from the supergravity 
and from the field-theory sides.

The main element of novelty that emerged from our study is that the dual field
theory undergoes two cascades: a cascade of Seiberg dualities, analogous
to the one in~\cite{KS}, and a cascade of Higgsings, analogous to what 
was suggested in~\cite{hep-th/0101013}. In both cases, the rank of the
gauge groups in the dual quiver field theory shrinks when going towards the IR.
But this is the result of very different phenomena.
The fact that the Higgsing cascade takes place over a compact,
 finite range of energy scales
is a significant element of novelty, also considering that traditional field theory studies
of this type of behavior are peculiarly difficult.

There are several open questions we leave for future studies.
On the more formal side, a well defined  project would be to determine from 
microscopic calculations the profiles (the function $S$ in the body of the paper) 
that reproduce the 
qualitative behavior we  proposed here. This would 
complete the picture in a satisfactory way.
Also, a  side project would be related to a precise discussion of the
existence (or not), couplings and properties of the 
Goldstone mode associated with the
baryonic symmetry, a problem that we only briefly touched upon in this work.

This work opens the way to several possible applications.
In particular, it would be interesting to make more concrete
the possible applications in the context of Dynamical Electroweak-Symmetry Breaking, for example by calculating  observables related to the  
discussion in Section \ref{SECTIONAPP}, or by adapting to  these backgrounds 
(or suitable modifications of them) the tests performed in \cite{pheno}. 
In this way, one might make some interesting progress towards 
the construction of a complete model of dynamical ElectroWeak Symmetry Breaking
in the rigorous context of gauge/string dualities, and hence move 
towards the direction of obtaining a fully computable and testable model of this type.

\section*{Acknowledgments}

Discussions with various colleagues
helped to improve the contents and  presentation of this paper.
We wish to thank Lilia Anguelova, Stefano Cremonesi, 
Anatoly Dymarsky (very specially), Daniel Elander, Aki Hashimoto, Tim Hollowood, Prem Kumar, Dario Martelli, Ioannis Papadimitriou and Jorge Russo.
The work of J.G. was funded by the DOE Grant DE-FG02-95ER40896. The  work of E.~C. and A.~V.~R.  was funded in part by MICINN   under grant
FPA2008-01838,  by the Spanish Consolider-Ingenio 2010 Programme CPAN (CSD2007-00042) and by Xunta de Galicia (Conseller\'\i a de Educaci\'on and grant INCITE09 206 121 PR) and by FEDER. E.~C is supported by a Spanish FPU fellowship, and thanks the FRont Of Galician-speaking Scientists for unconditional support.

\appendix
\renewcommand{\thesection}{\Alph{section}}
\renewcommand{\theequation}{\Alph{section}.\arabic{equation}}
\section{Technical aspects of the SUSY backgrounds without sources}
\label{backgroundfunctionsappendix}
\setcounter{equation}{0}
We write in this appendix various technical aspects
of the supersymmetric backgrounds without sources.
As explained in Section \ref{section2}, one can partially integrate the BPS system, and rewrite the functions $\Phi, h,g,k,a$ in terms of two new functions $P$ and $Q$. This was originally derived in
\cite{HoyosBadajoz:2008fw}, and the solution one finds is:
\begin{gather}
Q=N_c(2\r\,\coth(2\r)-1)\,, \quad e^{2h}\,=\,\frac{1}{4}\,
\frac{P^2-Q^2}{P\coth(2\r)-Q}\,,\quad e^{2g}\,=\,P\,\coth(2\r)\,-\,Q\,,\nonumber\\
e^{2k}=\frac{P'}{2}\,,\qquad a=\frac{P}{P\cosh(2\r)-Q\sinh(2\r)}\,,\qquad e^{4\Phi-4\Phi_0}= \frac{2\sinh(2\r)^2}{(P^2-Q^2)P'}\,.
\label{xxxxdd1.A}
\end{gather}
The function $b(\r)$ can be integrated exactly to give
\beq
b(\r)= \frac{2\r}{\sinh(2\r)}\,.
\label{bderho.A}\eeq
$P$ is given as the solution of the master equation (\ref{master}), that we recall here:
\beq
P'' + P'\Big(\frac{P'+Q'}{P-Q} +\frac{P'-Q'}{P+Q} - 4 
\coth(2\rho)\Big)=0\,.
\label{master.A}
\eeq
The solutions of the master equation are known only numerically, but their asymptotics can be written down analytically. In the UV (that is, large values of $\r$), we had the expansion (\ref{UV-II-N}):
\beq\bal
P=\,&e^{4\rho/3}\Big[ c_+  
+\frac{e^{-8\r/3} N_c^2}{c_+}\left(
4\r^2 - 4\r +\frac{13}{4} \right)+ e^{-4\r}\left(
c_- -\frac{8c_+}{3}\r \right)+\\
& + \frac{N_c^4 e^{-16\r/3}}{c_+^3}
\left(\frac{18567}{512}+\frac{2781}{32}\r +\frac{27}{4}\r^2 +36\r^3\right)  + \co(e^{-20\rho/3})
\Big]\,,
\label{UV-II-N.A}
\eal\eeq
which can be plugged back into eq.(\ref{xxxxdd1.A}) to obtain 
the background functions at large $\r$,
\begin{align}
& e^{2h}=\Big(\frac{c_+ e^{4\r/3}}{4}+\frac{N_c}{4}(2\r-1)+\frac{N_c^2 e^{-
4\r/3}}{16c_+}(16\r^2-16\r+13)+\cO(e^{-8\r/3})   \Big)\,,\NO\\
&\frac{e^{2g}}{4}=\Big(  
\frac{c_+ e^{4\r/3}}{4}-\frac{N_c}{4}(2\r-1)+\frac{N_c^2 e^{-
4\r/3}}{16c_+}(16\r^2-16\r+13)+\cO(e^{-8\r/3})\Big)\,,\NO\\
& \frac{e^{2k}}{4} = \Big(
\frac{c_+ e^{4\r/3}}{6}-\frac{N_c^2 e^{-4\r/3}}{24c_+}(4\r-5)^2+
\cO(e^{-8\r/3})   \Big)\,,\NO\\
& \frac{e^{4\Phi}}{e^{4\Phi(\infty)}}=1+\frac{3N_c^2 e^{-8\r/3}}{4c_+^2}(1-8\r)+
\frac{3 N_c^4 e^{-16\r/3}}{512 c_+^4}(2048\r^3+1152 \r^2 
+2352 \r -775)+\cO(e^{-8\r})\,,
\NO\\
&a= 2 e^{-2\r}+\frac{2N_c}{c_+}(2\r-1)e^{-10\r/3} 
+\frac{2N_c^2}{c_+^2}(2\r-1)^2 e^{-14\r/3}+\cO(e^{-9\r}) \,,       \NO\\
& b=\frac{2\r}{\sinh(2\r)}= 4\r\, e^{-2\r}+ 4\r\, e^{-6\r}+\cO(e^{-8\r})\,.
\label{gggg}
\end{align}
The geometry in eq.(\ref{f3old}) asymptotes to the conifold 
after using the expansions above.
In the IR (that is, close to the origin of the space that we take to be $\r=0$), we have to use eq.(\ref{P-IR}):
\beq
P= h_1 \r+ \frac{4 h_1}{15}\left(1-\frac{4 N_c^2}{h_1^2}\right)\r^3
+\frac{16 h_1}{525}\left(1-\frac{4N_c^2}{3h_1^2}-
\frac{32N_c^4}{3h_1^4}\right)\r^5+\co(\r^7)\,,
\label{P-IR.A}
\eeq
and (\ref{xxxxdd1.A}), to obtain
\begin{align}
&e^{2h}=\frac{h_1 \rho ^2}{2}+\frac{4}{45} \left(-6 h_1+
15 N_c-\frac{16 N_c^2}{h_1}\right) \rho ^4+\co(\r^6)\,,\NO\\
&\frac{e^{2g}}{4}=\frac{h_1}{8}+
\frac{1}{15} \left(3 h_1-5 N_c-\frac{2 N_c^2}{h_1}\right) \rho ^2+
\frac{2 \left(3 h_1^4+70 h_1^3 N_c-144 
h_1^2 N_c^2-32 N_c^4\right) \rho ^4}{1575 h_1^3}+\co(\r^6)\,,\NO\\
&\frac{e^{2k}}{4}=\frac{h_1}{8}
+\frac{\left(h_1^2-4 N_c^2\right) \rho ^2}{10 h_1}
+\frac{\left(6 h_1^4-8 h_1^2 N_c^2-64 N_c^4\right) \rho ^4}{315 h_1^3}
+\co(\r^6)\,,\NO\\
&e^{4(\Phi-\Phi_0)}=1+\frac{64  N_c^2 \rho ^2}{9 h_1^2}
+\frac{128 N_c^2 \left(-15 h_1^2+124 N_c^2\right) \rho ^4}{405 h_1^4}+
\co(\r^6)\,,\NO \\
&a = 1+\left(-2+\frac{8 N_c}{3 h_1}\right) \rho ^2
+\frac{2 \left(75 h_1^3-232 h_1^2 N_c+160 h_1 N_c^2+64 N_c^3\right) 
\rho ^4}{45 h_1^3}
+\co(\r^6)\,,\NO\\
&b = \frac{2\r}{\sinh(2\r)}= 1-\frac{2}{3}\r^2 
+\frac{14}{45}\r^4  +\co(\r^6)\, .
\label{SUSYIREXPapen}
\end{align}
This space is free of singularities as can be checked by computing curvature invariants.
\section{Detailed derivations in the case of sources with profile}
\setcounter{equation}{0}
\label{appendixdetailedderivation}
Let us work out the derivation of the supergravity solutions in Section \ref{AEJ}, for the case where we add sources with a profile $S(r)$. Suppose that we start with the master equation (\ref{master-eq}),
\beq
(P''+N_f S')+(P'+N_f S)\Big[\frac{P'+Q'+2 N_f S}{P-Q}
+ \frac{P'-Q'+2 N_f S}{P+Q}
- 4 \coth(2\r)   \Big]=0\,\,,
\eeq
that we will rewrite as
\beq
\partial_\r\Big( \frac{(P^2-Q^2)}{\sinh^2 (2\r)} (P'+N_f S) \Big) 
+\frac{4}{\sinh^2 (2\r)}(P'+N_f S)(QQ'+ PN_f S)=0\,\,.
\eeq
Then, we can take this last equation and integrate it twice, to get
\beq
	\bal
		&P^3 - 3 P Q^2 + 3 \int_0^\r d\tilde\r\, \big(2 PQQ' + N_f S(P^2-Q^2)\big)-\\
		&-12 \int_0^\r d\tilde{\r}\,\sinh^2(2\tilde{\rho})\int_{\tilde{\r}}^{\infty}d\hat{\r}\,\frac{(P'+N_f S)}{\sinh^2 2\hat\r}(QQ'+ PN_f S)=4\lambda^3 \e^4 \int_0^\r d\tilde\r \sinh^2(2\tilde\r)\,.
	\eal
\eeq
We will now propose a solution in an inverse series expansion in the constant 
$\lambda$
\beq
P= \lambda P_1 + P_0 + 
\frac{P_{-1}}{\lambda}+
\frac{P_{-2}}{\lambda^2}+
\frac{P_{-3}}{\lambda^3}+\ldots\,\,.
\eeq
By equating factors of $\lambda$ 
we get,
\beq
\bal
P_1&= \left(4\e^4 \int_0^\r d\tilde\r\,\sinh^2(2\tilde\r)\right)^{1/3}\,,\\
P_0&= -\frac{N_f}{P_1^2}\left(\int_0^\r d\tilde\r\, P_1^2 S - 4 \int_0^\r d\tilde\r\, \sinh^2(2\tilde\r)\int_{\tilde\r}^{\infty} d\hat\r \,
\frac{P_1 P_1' S}{\sinh^2(2\hat\r)} \right)\,,\\
P_{-1}&= -\frac{1}{P_1^2}\bigg(
P_1(P_0^2-Q^2)+2\int_0^\r d\tilde\r\, P_1(N_f P_0 S + QQ') -
\\
&\quad- 4 \int_0^\r d\tilde\r\, \sinh^2(2\tilde\r)\int_{\tilde\r}^{\infty} d\hat\r\, \frac{N_f P_1 S(N_f S + P_0')+P_1'(N_fP_0S 
+QQ')}{\sinh^2(2\hat\r)} \bigg)\,,\\
P_{-2}&= -\frac{1}{P_1^2}\bigg(\frac{P_0^3}{3} + 2 P_0 P_1 P_{-1}- P_0Q^2 +
\int_0^\r d\tilde\r\,\big(N_f P_0^2 S +N_fS(2 P_1P_{-1}-Q^2) +2 P_0 QQ'\big) -\\
&\quad- 4 \int_0^\r d\tilde\r\,\sinh^2(2\tilde\r)\int_{\tilde\r}^{\infty} d\hat\r\, \frac{N_f P'_1 P_{-1} S+N_f P_1 P'_{-1} S 
+(N_f S + P_0')(N_fP_0S +QQ')}{\sinh^2(2\hat\r)}\bigg)\,.
\eal
\label{lambdaseriesapp}
\eeq

\subsection{The limit $\l \to \infty$}
We describe in detail the limit of large $\l$. 
This will lead to an analytic solution.
As will see, we need to take at the same time 
$N_f\to 0$ to make sense of the background.
First of all, we write the explicit expansions for the 
functions in the limit
\beq
\l\to \infty\,,\qquad N_f\to 0\,,\qquad N_f \l\to \frac{2^{2/3}}{\e^{4/3}}\nu\,.
\label{limite}\eeq
They are
\begin{align}
e^{2h}=&\l \frac{\tanh(2\r)}{4}P_1(\r)
+\frac{\tanh(2\r)}{4}(P_0 + Q \tanh(2\r))+\nonumber\\
&+\frac{\tanh(2\r)}{4P_1 \l}(P_1 P_{-1}- \frac{Q^2}{\cosh^2(2\r)})+\co(\l^{-2})\,,\nonumber\\
\frac{e^{2g}}{4}=& \l \frac{\coth(2\r) P_1 }{4}+\frac{1}{4}(P_0 \coth(2\r) - Q) 
+\frac{P_{-1}\coth(2\r)}{4\l}+\co(\l^{-2})\,,\nonumber\\
\frac{e^{2k}}{4}=& \frac{P_1' \l}{8}+\frac{N_f S(\r) +P_0'}{8}+ 
\frac{P_{-1}'}{8\l}+ \co(\l^{-2})\,,\nonumber\\
a=&\frac{1}{\cosh 2\r}+\frac{Q\tanh(2\r)}{P_1 \cosh(2\r) \l}+\frac{Q\tanh(2\r)}{P_1\l^2}\left(Q\sinh(2\r)-P_0\cosh(2\r)\right)+\co(\l^{-3})\,,\nonumber\\
e^{4\Phi-4\Phi_0}=& \frac{2\sinh^2(2\r)}{\l^3 P_1^2 P_1'}-
\frac{2\sinh^2(2\r)}{\l^4 P_1^3 P_1'^2}(P_1(N_f S + P_0')+
2 P_0 P_1')+\nonumber\\
&+\frac{2\sinh(2\r)^2}{P_1^4P_1'^3\l^5}\left(P_1(P_0'+N_fS)(2P_0'P_1+2P_0P_1'+N_fP_1S)-P_0'^2P_1^2+\right.\nonumber\\
&\left.+P_1'(3P_0^2P_1'+P_1'Q^2-2P_1P_1'P_{-1}-P_1^2P_{-1}')\right)+\co(\l^{-6})\,,\nonumber\\
\hat{h}=& 1- \frac{\k^2 e^{2\Phi_0}}{\l^{3/2}}\sqrt{\frac{2\sinh^2(2\r)}{P_1^2 P_1'}} + 
\frac{\hat{h}_f}{\l^{5/2}}+ \frac{\hat{h}_c}{\l^{7/2}}+\co(\l^{-9/2})\,,
\label{functionslambdainfinityapp}
\end{align}
where we have defined
\beq
\bal
\hat{h}_f=&\frac{\k^2 e^{2\Phi_0}}{2P_1P_1'}\sqrt{\frac{2\sinh^2(2\r)}{P_1^2 P_1'}}(2P_0 P_1' +P_1(N_f S + P_0'))\,,\\
\hat{h}_c=&\frac{\k^2 e^{2\Phi_0}}{8 P_1'^2 P_1^2}\sqrt{\frac{2\sinh^2(2\r)}{P_1^2 P_1'}} \Big[-4 P_1'^2(2P_0^2+Q^2)  
+ 4 P_1P_1'(2P_{-1}P_1'- P_0P_0')+\\
&+P_1^2(4P_1'P_{-1}' -3 P_0'^2)
 -2 N_f P_1 S[3P_1 P_0' +2 P_0 P_1']- 3 N_f^2 P_1^2 S^2    \Big]\,.
\eal
\eeq
Notice that the expressions for $Q(\r),b(\r)$ do not change from those in
eqs.(\ref{Q-integral}) and (\ref{bderho}) respectively.
Now, we will consider the limit in eq.(\ref{limite}). 
We choose the value of the constant $\k$ as
\beq
3\k^4= 2\e^4 \l^{3}e^{-4\Phi_0}\,.
\eeq
Again the logic for this choice is to get a warp factor that vanishes at large $\r$.
In order to have a well-defined metric
(where powers of $\l$ are absent) when taking the limit 
$\l\to \infty$, we also need to take $N_f\to 0$ as in eq.(\ref{limite}).
Doing this has the following effect on the 
functions defining the full background:
\beq\begin{split}
&e^{2h}=\l \frac{\tanh(2\r)}{4}P_1(\r)+\cO(\l^0)\,,\quad
\frac{e^{2g}}{4}= \l \frac{\coth(2\r) P_1 }{4}+\cO(\l^0)\,,\\
&\frac{e^{2k}}{4}= \frac{P_1' \l}{8}+\cO(\l^0)\,,\quad
a= \frac{1}{\cosh 2\r}+\cO(\l^{-1})\,,\quad
e^{4\Phi}= \frac{3e^{4\Phi_0}}{2 \e^4 \l^3}+\cO(\l^{-4})\,,\\
&Q=N_c(2\r \coth(2\r)-1)+\cO(\l^{-1})\,,\quad
b=\frac{2\r}{\sinh(2\r)}+\cO(\l^{-1})\,,
\end{split}\eeq
and on the function $\hat{h}$ this has two interesting effects. On the one hand
taking $N_f\to 0$ considerably simplifies 
the expression for $\hat{h}_c,\hat{h}_f$. On the other hand the scaling 
$N_f\l=2^{2/3} \e^{-4/3} \nu$ fixed 
makes the term $\l^{-5/2} \hat{h}_f$ scale like $\l^{-7/2} \hat{h}_c$
(that is as $\l^{-2}$).
Notice also that in the limit of eq.(\ref{limite}), $\l^{-5/2} \hat{h}_f$ becomes
\beq
\l^{-5/2} \hat{h}_f= \l^{-2} \frac{4}{\e^{8/3}} \nu\int_{\r}^{\infty} 
dx \frac{S(x)}{(\sinh(4x)-4x)^{1/3}}\,.
\label{h1app}
\eeq
Regarding the integrals defining $\hat{h}_c$, 
we see that in the limit of eq.(\ref{limite}), we
have
\beq
\frac{\hat{h}_c}{\l^{7/2} } = \frac{2^{5/3}}{\l^2\e^{8/3}} N_c^2 \int_{\r}^{\infty} dx\,
\frac{(\sinh(4x)-4x)^{1/3}(2x \coth(2x)-1)}
{\sinh^2(2x)} +\cO(\l^{-3})=\frac{\hat{h}_{KS}}{\l^2}+\cO(\l^{-3})\,.
\label{h2app}
\eeq
Notice that we dropped terms
that are suppressed like $\frac{\nu^2}{\l^2}$ in the previous expression.
Finally, rescaling the Minkowski coordinates
$x_1\to x_i \l^{-1/2}$ we have a metric that is independent of the parameter
$\l$. Using that
\beq
\frac{P_1'}{P_1}=\frac{8 \sinh^2(2\r)}{3(\sinh(4\r)-4\r)}\,\,,
\eeq
the internal space metric is the deformed conifold.
In this way, we have generated a new analytic family of solutions
that depend on the function $S(\r)$.
For correct choices of $S(\r)$, like for 
example $S(\r)\sim (\tanh(2\r))^{2n}$
as in 
\cite{Barranco:2011vt} or those in eq.(\ref{vvvvzzzz}), 
this family of solutions is non-singular.
Interestingly, it is also true when we do not take the limit described in 
eq.(\ref{limite}).
In that case, our series expansion for $P(\r)$ and $\hat{h}$ are not truncated.
If we take the strict $N_f=0$ case, this solution is the baryonic branch of
\cite{BGMPZ}. The parameter $\l$
is the one moving between different VEV's for the baryon and anti-baryon 
operators
(a parameter called ${\cal U}$ in  \cite{Dymarsky:2005xt}).
For nonzero values of $N_f$, an interpretation
of the parameter $\l$ was given in \cite{GMNP}.
\subsection{The exact and analytic solution}
We describe here the solution obtained in the limit $\lambda\to \infty$.
The warp factor is
\beq
\hat{h}= \frac{1}{\lambda^2} \left( \frac{4}{\e^{8/3}}
\nu \int_\r^{\infty} dx\frac{S(x)}{(\sinh(4x)-4x)^{1/3}} + \hat{h}_{KS}\right)\,.
\eeq
After rescaling $x_i\to x_i \l^{-1}$ and choosing $\Phi(\infty) = 0$,
the 
full configuration will read
\beq
\bal
ds^2&= \hat{h}^{-1/2}dx_{1,3}^2 + \hat{h}^{1/2}ds_6^2\,,\\
ds_6^2&=P_1 \Big(\frac{P_1'}{2P_1}
(d\r^2+\frac{1}{4}(\tilde{\omega}_3 +\cos\theta d\varphi)^2) + 
\frac{\tanh(2\r)}{4}(d\theta^2+\sin^2\theta d\varphi^2)+\\
&\qquad\quad+\coth(2\r)\left[(\tilde{\omega}_1+a d\theta)^2+(\tilde{\omega}_1-
a \sin\theta d\varphi)^2\right]   \Big)\,,\\
B_2&= - \frac{N_c}{4} \frac{2\r \coth(2\r) -\!1}{\sinh(2\r)}\!\left[\cosh(2\r) (\sin \theta d\theta \wg d\varphi - \sin \tilde{\theta} d\tilde\theta \wg d \tilde \varphi)\!-\sin \theta d\varphi \wg \tilde \w_1 - d \theta \wg \tilde\omega_2 \right]\!,\\ 
H_3 &= dB_2\,,\qquad\quad F_3=*_6 H_3\,,\\
F_5&=-(1+*)\partial_\r H^{-1} dt\wedge dx_1\wedge dx_2\wedge dx_3 \
\wedge d\r\,,\\
H&= \hat{h}_{KS} + \frac{4}{\e^{8/3}} \nu \int_{\r}^{\infty}dx 
\frac{S(x)}{(\sinh(4x)-4x)^{1/3}}\,.
\eal
\label{exactsolutionapp}
\eeq
Here, taking the limit for $B_2$ is non-trivial. Indeed, when expanding it in terms of inverse powers of $\l$, one gets
\beq
	B_2 = \l \frac{\e^2 \sinh(2\r)}{2\sqrt{3} \k P_1\sqrt{P_1'}} \,d\Big(P_1(\tilde\omega_3 +\cos\theta\, d\varphi)\Big) - B_{KS} + \cO\left(\l^{-1}\right)\,.
\eeq
Using eq.\eqref{P1}, one notices that the overall factor in front of the total derivative in the previous expression is a constant. So the leading order term in $B_2$ can be gauged away because it is exact. So one is left with $B_2 = - B_{KS} + \cO\left(\l^{-1}\right)$, which has a well-defined limit when $\l \to \infty$. Note that the sign difference between our result and the one of Klebanov-Strassler comes from a different choice of orientation.

\section{Equations of motion}
\label{app:EOM}

In this appendix  we state the equations of motion of the problem studied in this paper, that is type IIB supergravity with brane sources. The interested reader will find more details about the action from which those equations are derived in \cite{GMNP}.
Let us first start by writing the Bianchi identities, which are modified due to the presence of sources. From equations \eqref{configurationfinal} and \eqref{ffs}, one can see that the three-form flux $F_3$ is not closed anymore: 
\beq\bal
d F_3 =&\frac{N_f}{4}\,\sin\theta\,d\theta\wedge d\phi\wedge
\Big[\,S\,\tilde\omega^1\wedge\tilde\omega^2\,-\,S'\,d \r\wedge \tilde\omega^3\,\Big]
\,+\\
&+\frac{N_f}{8}\, \frac{S'}{\cosh (2\r)}\,
d \r\wedge \Bigg[\,d\theta\wedge 
\tilde\omega^2\wedge\tilde\omega^3+d\phi\wedge \Big(
\sin\theta \,\tilde\omega^1\wedge\tilde\omega^3+\cos\theta \,d\theta\wedge 
\tilde\omega^2\Big)\Bigg]\,\,.
\eal
\label{Omega-S(r)}
\eeq
We then define a four-form $\X_4$ as
\beq
\X_4=d F_3\,.
\eeq
That means that we get the following Bianchi identities for the other fluxes:
\beq
	\begin{aligned}
		d H_3 &= 0\,, &\qquad d F_{5} &= H_3 \wedge F_{3} + B_2 \wedge \X_{4}\,.
	\end{aligned}
\eeq
The equations of motion for the fluxes read
\beq
	\begin{aligned}
		d \left( e^{-\F} * H_3 \right) &= F_{3} \wedge F_{5} + e^{\F} \frac{\sqrt{1-\hat{h}}}{\hat{h}} \,\vol_{4} \wedge \X_{4}\,, \\
		d \left( e^{\F} * F_{3} \right) &= - H_3 \wedge F_{5}\,,
	\end{aligned}
\eeq
where $\vol_{4}=dx^0\wedge dx^1\wedge dx^2 \wedge dx^3$. We then define the following notation
\beq
	\o_{(p)} \lrcorner \l_{(p)} = \frac{1}{p!} \o^{\mu_1 ... \mu_p} \l_{\mu_1 ... \mu_p}\,.
\eeq
We also have that
\beq
	\int \o_{(p)} \wedge \l_{(10-p)} = - \int \sqrt{-g} \l \lrcorner (*\o)\,.
\eeq
Using these, we can write the dilaton equation of motion as
\beq
	\frac{1}{\sqrt{-g}} \partial_{\mu} \left( \sqrt{-g} g^{\mu \nu} \partial_{\nu} \F \right) = 
\frac{1}{12} e^{\F} F_{3}^2 - \frac{1}{12} e^{-\F} H_3^2 - \frac{1}{2} e^{\F/2} \X_{4} \lrcorner * \left( \frac{e^{3\F/2}}{\hat h} \vol_{4} \wedge J \right)\,,
\eeq
where $J$ is the almost-K\"ahler form of the internal space and is equal to
\beq
	J = \hat{h}^{1/2} e^{-\F/2} \Big[e^{\r 3}-\cos\alpha (e^{\theta\varphi}+ e^{12})-\sin\alpha(e^{\theta2}+ e^{\varphi 1}) \Big]\,,
\eeq
using the conventions of equations \eqref{vielbeinafter}-\eqref{cosinh}.
Finally, the Einstein equation is
\beq \label{eq:EinsteinEquation}
\bal
	R_{\m \n} -\frac{1}{2} g_{\m\n} R = &\frac{1}{2} \partial_{\m} \F \partial_{\n} \F - \frac{1}{4} g_{\m\n} \partial_{\s} \F \partial_{\s} \F + \frac{1}{24} e^{\F} \left( 6F_{\m \t \s} F_{\n}^{\phantom{\n} \t \s} - g_{\m \n} F_{3}^2 \right) +\\
	&+ \frac{1}{24} e^{-\F} \left( 6 H_{\m \t \s} H_{\n}^{\phantom{\n} \t \s} - g_{\m \n} H_3^2 \right)+ \frac{1}{96} F_{\m \t_1 \t_2 \t_3 \t_4} F_{\n}^{\phantom{\n} \t_1 \t_2 \t_3 \t_4} +T_{\m\n}^{\text{sources}}\,,
\eal
\eeq
where $T_{\m\n}^{\text{sources}}$ is the energy-momentum tensor coming from the source action. It is given by:
\beq \label{eq:EnergyMomentumTensor}
\bal
	T_{\m\n}^{\text{sources}} = &- \frac{1}{12} \frac{e^{2\F}}{\hat h} \left( \X_{\m \t_1 \t_2 \t_3} * \left( \vol_{4} \wedge J \right)_{\n}^{\phantom{\n} \t_1 \t_2 \t_3} -6 g_{\m \n} \X_{4} \lrcorner * \left( \vol_{4} \wedge J \right) \right)- \\
		&- \frac{1}{240} e^{\F} \frac{\sqrt{1-\hat h}}{\hat h}\left( (B_2 \wedge \X_{4})_{\m \t_1 ... \t_5}  \left( * \vol_{4} \right)_{\n}^{\phantom{\n} \t_1 ... \t_5} - 120 g_{\m \n} (B_2 \wedge \X_{4}) \lrcorner \left( * \vol_{4} \right) \right)\,.
\eal
\eeq
From \eqref{eq:EinsteinEquation} and \eqref{eq:EnergyMomentumTensor}, one can get an equation for the Ricci tensor as:
\beq
	\begin{aligned}
		R_{\m \n} = &\frac{1}{2} \partial_{\m} \F \partial_{\n} \F + \frac{1}{48} e^{\F} \left( 12 F_{\m \t \s} F_{\n}^{\phantom{\n} \t \s} - g_{\m \n} F_{3}^2 \right) +	\frac{1}{48} e^{-\F} \left( 12 H_{\m \t \s} H_{\n}^{\phantom{\n} \t \s} - g_{\m \n} H_3^2 \right)+ \\
		&+ \frac{1}{96} F_{\m \t_1 \t_2 \t_3 \t_4} F_{\n}^{\phantom{\n} \t_1 \t_2 \t_3 \t_4}- \\
		&- \frac{1}{24} \frac{e^{2\F}}{\hat h} \left( 2 \X_{\m \t_1 \t_2 \t_3} * \left( \vol_{4} \wedge J \right)_{\n}^{\phantom{\n} \t_1 \t_2 \t_3} -3 g_{\m \n} \X_{4} \lrcorner * \left( \vol_{4} \wedge J \right) \right)- \\
		&- \frac{1}{240} e^{\F} \frac{\sqrt{1-\hat h}}{\hat h}\left( (B_2 \wedge \X_{4})_{\m \t_1 ... \t_5}  \left( * \vol_{4} \right)_{\n}^{\phantom{\n} \t_1 ... \t_5} - 60 g_{\m \n} (B_2 \wedge \X_{4}) \lrcorner \left( * \vol_{4} \right) \right)\,.
	\end{aligned}
\eeq

\begin{figure}[t]
\begin{center}
\includegraphics[width=0.6\textwidth]{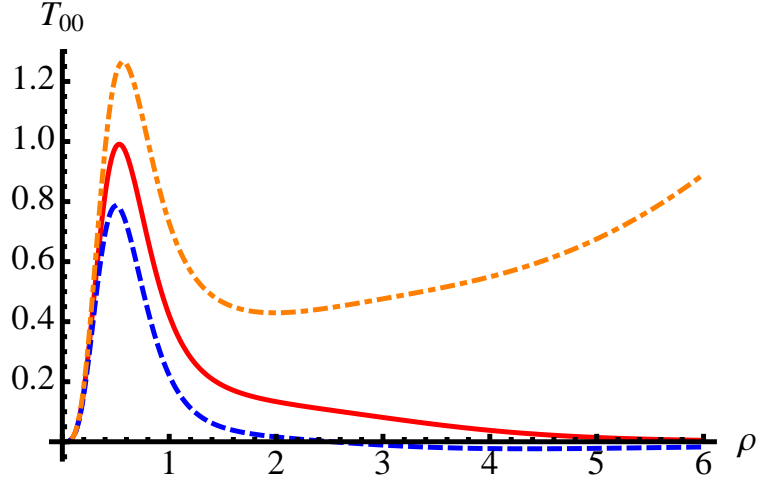}
\caption{Plots of $T_{{x^0\,x^0}}^{\text{sources}}$ (for $N_c=2=N_f$ and $h_1=5$) for different profiles. The dot-dashed orange curve corresponds to a profile $S$ decreasing like $e^{-(4-\epsilon)\r/3}$. It grows exponentially in the UV. The red curve is for the profile in eq.\eqref{vvvvzzzz}, and it behaves nicely. And the dashed blue curve corresponds to a profile $S$ decreasing like $e^{-(4+\epsilon)\r/3}$. It becomes negative for large values of $\r$. We take $\epsilon=1$ to appreciate well the qualitative difference, but the results hold for any $\epsilon>0$.}
\label{fig:T00}
\end{center}
\end{figure}

To clarify the contribution of the sources to the problem, let us express in more details the various components of the energy-momentum tensor due to the sources. In flat indices, the energy-momentum tensor of the sources gives:
\beq
	\begin{aligned}
		T_{{x^i\,x^j}}^{\text{sources}} &= -\frac{N_f}{2\hat h^{3/2}}\, e^{-2g-2h-2k-\Phi/2}  \left( e^{2k} S + \frac{4e^{2h} +e^{2g} \left(a-\cosh(2\r)\right)^2}{\sinh(4\r)}S' \right)\eta_{ij}\,,\\
		T_{{\r\r}}^{\text{sources}} &= -\frac{N_f}{2\hat h^{1/2}}\, e^{-2g-2h-\Phi/2} S = T_{{33}}^{\text{sources}}\,, \\
		T_{{\theta\theta}}^{\text{sources}} &= -\frac{N_f}{\hat h^{1/2}\sinh(4\r)}\, e^{-2g-2k-\Phi/2} S' = T_{{\phi\phi}}^{\text{sources}}\,, \\
		T_{{11}}^{\text{sources}} &= -\frac{N_f}{2\hat h^{1/2}}\, e^{-2g-2h-2k-\Phi/2} \, \frac{2e^{2h} +e^{2g} \left(a-\cosh(2\r)\right)^2}{\sinh(4\r)}S'  = T_{{22}}^{\text{sources}}\,, \\
		T_{{\theta1}}^{\text{sources}} &= \frac{N_f}{2\hat h^{1/2}}\, e^{-g-h-2k-\Phi/2}\, \frac{a-\cosh(2\r)}{\sinh(4\r)} S' = T_{{\phi2}}^{\text{sources}}\,.
	\end{aligned}
	\label{emtensor-components}
\eeq
where $\eta_{ij}$ is the flat Minkowski metric. Notice that this energy-momentum tensor is the same one (up to global factors of $\hat h$) as the one in the unrotated background \cite{Conde:2011rg}: the rotation does not change it.

One expects that the $tt$-component of the energy-momentum tensor, $T_{{x^0\,x^0}}^{\text{sources}}$, represents the mass density of sources in our `static' background. As such, it should be a positive-definite quantity (even if we perform a Lorentz transformation, its sign should be preserved). If we look at its expression from eq.\eqref{emtensor-components}, we have that
\beq
T_{{x^0\,x^0}}^{\text{sources}}=\frac{N_f}{2\hat h^{3/2}}\, e^{-2g-2h-2k-\Phi/2}  \left( e^{2k} S + \frac{4e^{2h} +e^{2g} \left(a-\cosh(2\r)\right)^2}{\sinh(4\r)}S' \right)\,.
\eeq
It is trivial that this positivity condition will hold for profiles like the one in Figure \ref{fig:D5s}, as both $S$ and $S'$ are positive. However, for profiles like eq.\eqref{vvvvzzzz} that we are studying in this paper, this condition is not trivially satisfied anymore.

We have seen (see Figure \ref{fig:T00}) that $S$ should not decrease faster than $e^{-4\r/3}$ in order for $T_{{x^0\,x^0}}^{\text{sources}}$ to be positive everywhere. Moreover, we also know that $S$ should decrease at least as fast as $e^{-4\r/3}$ in order to preserve the KS UV asymptotics. 
This argument seems to imply that the only physical choice is actually the one we have made in eq.\eqref{vvvvzzzz}. Notice that this choice is the only one that naturally gives a constant density of D3 sources in the rotated background, in virtue of eq.\eqref{distributionxxx}.

\end{document}